\documentclass[amsmath,amssymb,preprint]{revtex4}
\usepackage{amsmath}
\usepackage{graphicx}
\begin{document}

 \title{Soft disks in a narrow channel}

 \author{D. Mukamel}
 \email{david.mukamel@weizmann.ac.il}
 \affiliation{Department of physics of complex systems,
 Weizmann Institute of Science\\
 Rehovot, Israel 76100}

 \author{H. A. Posch}
 \email{Harald.Posch@univie.ac.at}
 \affiliation{Computational Physics Group, Faculty of Physics,
 Universit\"at Wien,
 Boltzmanngasse 5, A-1090 Wien, Austria}

 \date{\today}

\begin{abstract}
The pressure components of "soft" disks in a two dimensional narrow
channel are analyzed in the dilute gas regime using the Mayer
cluster expansion and molecular dynamics. Channels with either
periodic or reflecting boundaries are considered. It is found that
when the two-body potential, $u(r)$, is singular at some distance
$r_0$, the dependence of the pressure components on the channel width
exhibits a singularity at one or more channel widths which are
simply related to $r_0$. In channels with periodic boundary
conditions and for potentials which are discontinuous at $r_0$, the
transverse and longitudinal pressure components exhibit a $1/2$ and
$3/2$ singularity, respectively. Continuous potentials with a power
law singularity result in weaker singularities of the pressure
components. In channels with reflecting boundary conditions the
singularities are found to be weaker than those corresponding to
periodic boundaries.
\end{abstract}

 \maketitle
\section{Introduction}
\label{Motivation}

The thermodynamic and dynamical properties of particles in
restricted geometries are of great interest. They have been
extensively studied in the context of porous media
\cite{Given,Konig,Evans,Smit}, transport through narrow channels
such as carbon nanotubes \cite{Hummer,Allen} and pores in biological
membranes \cite{Hille} as well as in numerous other systems
\cite{Gelb}. Perhaps the simplest and most convenient theoretical
approach for studying fluids in cavities models the fluid by hard
spheres. This approach has been applied in a large number of studies
(see for example, \cite{Kamenetskiy,Kim,Mon2000,Kofke,Klafter}).
Recent studies of a dilute gas of hard disks in a narrow two
dimensional channel have shown that the system exhibits a
singularity in the pressure at a channel width equal to twice the
diameter of the disks \cite{FMP04,Bowles}. This is a consequence of the
fact that the volume of the phase space available to the disks
exhibits a singularity at this width. In particular, it has been
found that for a channel with periodic boundary condition, the
transverse component of the pressure exhibits a $3/2$ singularity,
while the longitudinal component exhibits a $5/2$ singularity
\cite{FMP04}. In the case of a channel with reflecting boundaries,
weaker singularities for both pressure components were found
\cite{Bowles}.

In this paper we extend these studies to consider a gas of "soft"
disks in a narrow channel at low density. We consider several
classes of two body potentials, with both periodic and reflecting
channel boundaries. Our analysis shows that the pressure components
are singular at some channel widths whenever the interaction
potential between two disks, $u(r)$, is singular at some distance
$r_0$. In particular, in the case of periodic boundary conditions and
for potentials which are discontinuous at some $r_0$, the dependence
of the transverse component of the pressure on the channel width
exhibits a $1/2$ singularity, while the longitudinal component
exhibits a $3/2$ singularity at some channel widths, which are
simply related to $r_0$. The nature of these singularities becomes
weaker for interaction potentials which are continuous, but they still display
a power law singularity at $r_0$. The singularities of the case of
reflecting boundary conditions are found to be weaker than those
corresponding to periodic boundaries. Although we have not analyzed
narrow channels in three dimensions, we expect similar phenomena to
take place there as well. The nature of the singularities in the
pressure in three dimensions is expected to be different from that
of the two dimensional case.

In the following sections we study several classes of two body
potentials $u(r)$ for both periodic and reflecting boundary
conditions. In Section \ref{formulation} we present the general
formulation of the tools used in this study, the Mayer cluster
expansion for gases at low densities, and molecular dynamics. In
Section \ref{Periodic-boundaries} we analyze the case of a channel
with periodic boundary conditions, for one-step and two-steps
potentials. We also study a smooth potential which vanishes with a
power law at a critical distance. In Section
\ref{reflecting-boundaries} we study a channel with reflecting
boundaries for the cases of soft disks and soft disks with a hard
core. Finally, a brief summary is given in Section
\ref{conclusions}.

\section{General Formulation}
\label{formulation}

We consider $N$ disks of diameter $d$ and mass $m$, interacting via
a two body potential $u(r)$ at temperature $T$. The disks are
restricted to move in a channel of length $L_x$ and width $L_y$ with
$L_y \ll L_x$. In this study we analyze the pressure components of
this gas using virial expansion to second order in the density. We
also carry out molecular dynamics simulations of this system. In the
present study the channel width, $L_y$, is taken to be finite
throughout the calculation. Therefore the free energy is not
extensive in $L_y$ and, thus, Euler's relation does not hold, namely,
$-PV \neq E-TS-\mu N$. Thus, the pressure has to be calculated by
taking the appropriate derivative of the free energy.

In the grand canonical ensemble the free energy of a system of $N$
disks is given by
\begin{eqnarray}
F(T,V,N) &=& -kT\ln{{\mathcal L}(T,V,z)} + kTN\ln{z}\nonumber \;,\\
N &=& z{\frac{\partial}{\partial z}}\ln{{\mathcal L}(T,V,z)} \;.
\label{free-energy}
\end{eqnarray}
Here ${\mathcal L}$ is the grand partition sum, $z$ is the
fugacity, $k$ is the Boltzmann constant and $V=L_xL_y$. The
pressure components are evaluated by taking the appropriate
derivatives of the free energy
\begin{eqnarray}
P_{xx}V &=&-L_x {\frac{\partial F}{\partial L_x}} \nonumber \\
P_{yy}V &=&-L_y {\frac{\partial F}{\partial L_y}} \nonumber \\
P &=& {\frac{1}{2}}(P_{xx} + P_{yy}) \;.
\label{pressure-components}
\end{eqnarray}

To second order in the fugacity $z$, the Mayer expansion yields
\begin{eqnarray}
\ln{\mathcal L} &=& {\frac{V}{\lambda^2}}(b_1 z + b_2 z^2)\nonumber\\
{\frac{1}{v}}\equiv {\frac{N}{V}} &=&  {\frac{1}{\lambda^2}}(b_1 z
+ 2b_2 z^2)\;, \label{Mayer-expansion}
\end{eqnarray}
where $\lambda=h/\sqrt{2 \pi mkT}$ is the average thermal
wavelength and $h$ is Planck's constant. The coefficients of the
expansion satisfy
\begin{eqnarray}
b_1 &=& 1\nonumber \\
b_2 &=& {\frac{1}{2\lambda^2}} q(L_y) \;,
\end{eqnarray}
with
\begin{equation}
q(L_y)=\int f_{12} d^2{r_{12}} ~ .
\label{q}
\end{equation}
Here, $f_{12} = e^{-\beta u(r_{12})}-1$ is the Mayer function, and $\beta = 1/kT$. Using
Eqs. (\ref{free-energy},\ref{Mayer-expansion}) we find that to order
$1/v$ the free energy is given by
\begin{equation}
\frac{F}{kTN} = -1-{\frac{q(L_y)}{2v}} -\ln{v} +2\ln{\lambda}
\quad ,
\end{equation}
from which the components of the pressure tensor are obtained:
\begin{eqnarray}
\frac {P_{xx}v}{kT} &=& 1 - \frac{q{(L_y)}}{2v} \,,\label{pxxgeneral}\\
\frac {P_{yy}v}{kT} &=& 1 - \frac{q{(L_y)}}{2v}+\frac{L_y}{2v} \frac{d q(L_y)}{d L_y} \;.
\label{pyygeneral}
\end{eqnarray}

All theoretical considerations are augmented by computer
simulations.  For our two-dimensional systems, the
temperature $T$ is computed from
\begin{equation}
    \langle K \rangle = \left(N - \frac{g}{2}\right) kT,
   \label{kinetic}
\end{equation}
where $K$ is the kinetic energy and the bracket denotes a time
average. Here, $g$ is  the number of  macroscopic conservation laws,
which differs for the periodic boundaries of Section
\ref{Periodic-boundaries}, $(g = 3$; the total momentum is constant and is taken to vanish),
and for the reflecting boundaries used in Section {\ref{reflecting-boundaries} $(g = 1)$.
The diagonal elements of the
pressure tensor, $P_{\alpha \alpha}$ for  $\alpha \in \{ x,y\}$, are
evaluated from the  virial theorem,%
\begin{equation}
P_{\alpha \alpha}  V = \langle K \rangle +  W_{\alpha \alpha}  .
\label{virialtheorem}
\end{equation}
For impulsive interactions \cite{Rapaport}, the potential contribution, $W_{\alpha \alpha}$,
is given by
\begin{equation}
 W_{\alpha \alpha} = \frac{1}{ \tau } \sum_{c}
r_{\alpha,ij}^{(c)} \Delta v_{\alpha,i}^{(c)} \; ,
\label{VTcoll}
\end{equation}
Here, the sum is over all collisional events $c$, which instantaneously change the
potential energy during the averaging time $\tau$,
$ r_{\alpha,ij}^{(c)} \equiv r_{\alpha,i}^{(c)} - r_{\alpha,j}^{(c)}$ is the
$\alpha$-component of the separation vector of the two particles involved in the event,
and $\Delta v_{\alpha,i}^{(c)}$ denotes the velocity
change  for particle $i$ parallel to $\alpha$ due to that event
(the velocity change  for particle $j$ being just the opposite).
For the continuous potentials of  Section \ref{PLP}, the potential contribution becomes
\begin{equation}
 W_{\alpha \alpha} = \frac{1}{\tau} \int_0^{\tau} d t \sum_i\sum_{j>i} r_{\alpha,ij}
 f_{\alpha,ij} \;,
 \label{VTcont}
\end{equation}
where $f_{\alpha,ij}$ is the $\alpha$-component of the force exerted on $i$ by particle $j$.
In all figures comparing experimental with theoretical pressures, the experimental dots
represent   $W_{\alpha \alpha} / \langle K \rangle$.
 The theoretical smooth curves represent  $(P_{\alpha \alpha} v / k T) - 1$,
where the  temperature required for this computation is taken from
Eq. (\ref{kinetic}). As usual, temperature units are used, for which Boltzmann's constant $k$
is unity. Because of the equivalence of the canonical and microcanonical
ensembles, the experimental and theoretical pressures should agree up to a  term
of order O$(1/N)$. To make this correction insignificant,  at least 60 particles, or even
120 in many cases, were used for the simulations.

An event-driven algorithm is used \cite{Rapaport,AT} for the discontinuous
potentials with periodic  (Sec. \ref{Periodic-boundaries}) or reflecting
(Sec. \ref{reflecting-boundaries}) boundary conditions,
for which instantaneous  potential energy changes and boundary crossings
of a particle  are considered as events. For the continuous power-law potentials
of Sec. \ref{PLP} a hybrid code is used, which will be described there in more detail.

\section{Narrow channels with periodic boundary conditions}
\label{Periodic-boundaries}

\subsection{Positive step potential}

We proceed by considering the pressure in the case of a
two-body step potential

\begin{equation}
u(r)= \left\{ \begin{array}{ll} u & \qquad r \leq d \\
0 & \qquad r>d \,,
\end{array} \right.
\label{potential1}
\end{equation}
where $u>0$ is a constant (in our previous paper \cite{FMP04} we
considered the hard-disk case $u=\infty$). The case $u<0$ is
pathological, since the particles may collapse to form a cluster, as
long as there is no repulsive interaction at short distances. The
case of a two step potential with repulsion at short distances  and
attraction at larger distances will be considered in the following
sub-section. Here we limit ourselves to the repulsive potential case
$u>0$. To evaluate the pressure we associate with each particle an
interaction disk of radius $d$ centered at its position. Two
particles $i$ and $j$ interact with each other, if the center of $j$
is within the interaction disk of $i$, and {\em vice versa}. In the
case $L_y > 2d$ the cross sections of a particle (a disk of radius
d) and that of its image resulting from the periodic boundary
conditions in the $y$ direction do not overlap. Hence the integral
(\ref{q}) simply yields
\begin{equation}
q(L_y)=\pi d^2 (e^{-\beta u}-1) \qquad \mbox{for} \qquad L_y>2d \,.
\label{qwide}
\end{equation}
For $L_y <2d$ we note that the area of overlap between the
interaction disk of a particle and that of its image translated in
the $y$ direction is given by
\begin{equation}
S(\vartheta)=d^2(\pi -2 \vartheta - \sin{2 \vartheta})\:,
\label{area}
\end{equation}
where $\vartheta$ satisfies (see Fig.~(\ref{Fig1}))
\begin{equation}
L_y=2d\sin\vartheta\,.
\label{L_y}
\end{equation}
In this case the integral (\ref{q}) yields
\begin{equation}
q(L_y)=(\pi d^2 - 2 S(e^{-\beta u}-1)+ S(e^{-2\beta u}-1)\;.
\label{qnarrow}
\end{equation}
%
%
\begin{figure}
\centering
{\includegraphics[width=8cm]{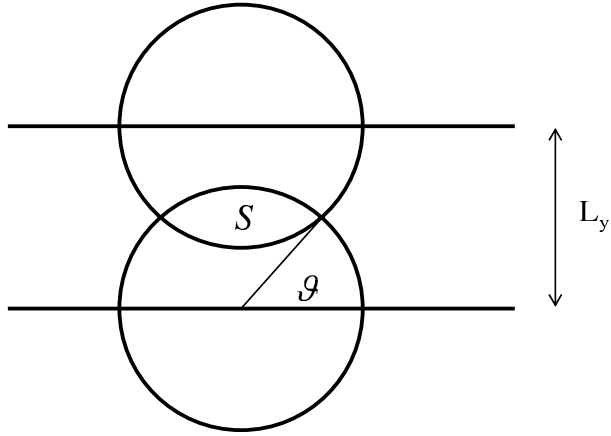}}
\caption{The interaction disks arrangement for $L_y < 2d$ with periodic
boundary conditions. The overlap area of the two disks is $S$.}
\label{Fig1}
\end{figure}
Using this result for $q(L_y)$, and noting that
\begin{equation}
\frac{dS}{dL_y}=-\sqrt{4d^2-L_y^2}~\quad ,
\end{equation}
it is straightforward to derive the expressions for the pressure. We
find that to order $1/v$ and for $L_y>2d$ one has $P_{xx}=P_{yy}=P$
with
\begin{equation}
\frac{Pv}{kT}=1 - \frac{q}{2v}\,,
\end{equation}
where $q$ is independent of $L_y$ and is given by (\ref{qwide}). On
the other hand, for $L_y<2d$ one finds
\begin{eqnarray}
\frac {P_{xx}v}{kT} &=& 1 - \frac{q{(L_y)}}{2v} \,,\label{pxxwide}\\
\frac {P_{yy}v}{kT} &=& 1 - \frac{q{(L_y)}}{2v}-\frac{L_y}{2v}\sqrt{4d^2-L_y^2}\left(1-e^{-\beta u}\right)^2 \;,
\label{pyywide}\\
\frac{Pv}{kT}     &=& 1 - \frac{q{(L_y)}}{2v}-\frac{L_y}{4v}\sqrt{4d^2-L_y^2}\left(1-e^{-\beta u}\right)^2 \:,
\label{pwide}\
\end{eqnarray}
where $q(L_y)$ is given by (\ref{qnarrow}). It is evident that
$P_{yy}$ exhibits a square-root singularity at $L_y=2d$ as in the
case of hard disks \cite{FMP04}. This singularity originates from
the term $dq(L_y)/dL_y$ in (\ref{pyygeneral}). On the other hand,
$P_{xx}$ exhibits a weaker singularity with a singular term which
vanishes as $(2d-L_y)^{3/2}$. The reason is that unlike the $P_{yy}$
component, here the singularity originates from $q(L_y)$ and not
from its derivative. Clearly, the pressure $P$, which is the average
of the two components, exhibits a square-root singularity as the
more singular $P_{yy}$ component.

\begin{figure}
\centering
{\includegraphics[width=10cm]{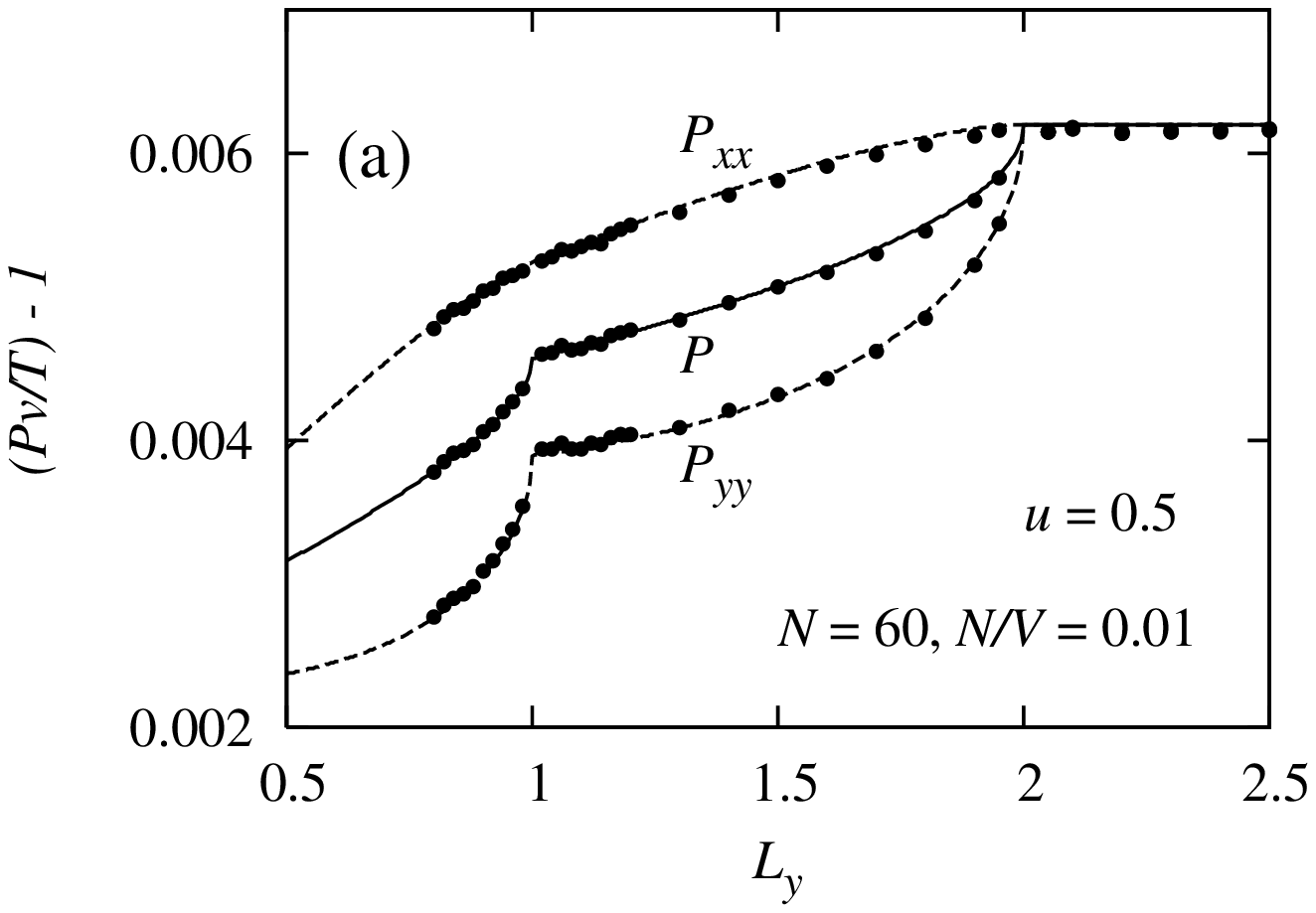}\\
\vspace{-5mm}
 \includegraphics[width=10cm]{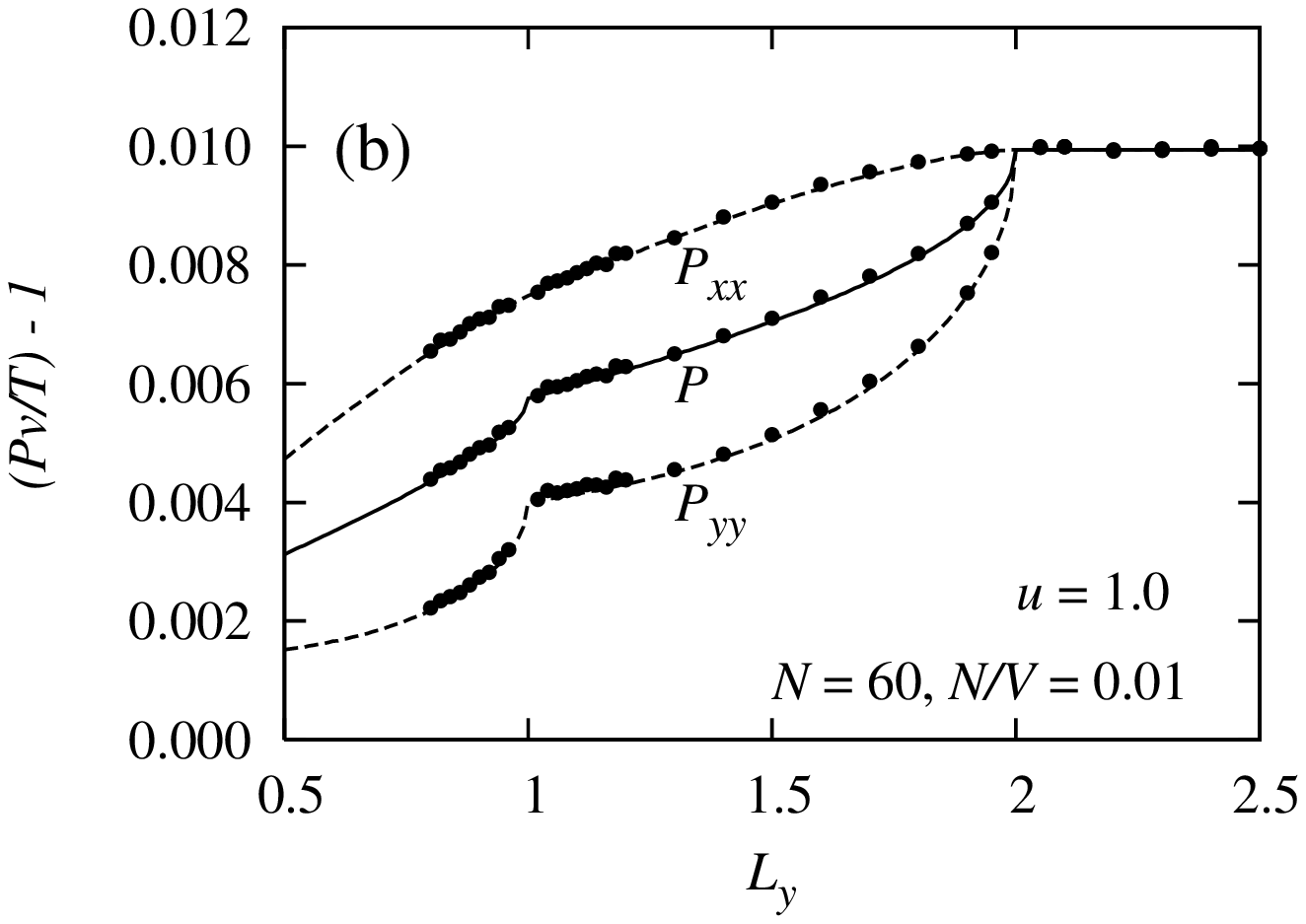}\\
\vspace{-5mm}
 \includegraphics[width=10cm]{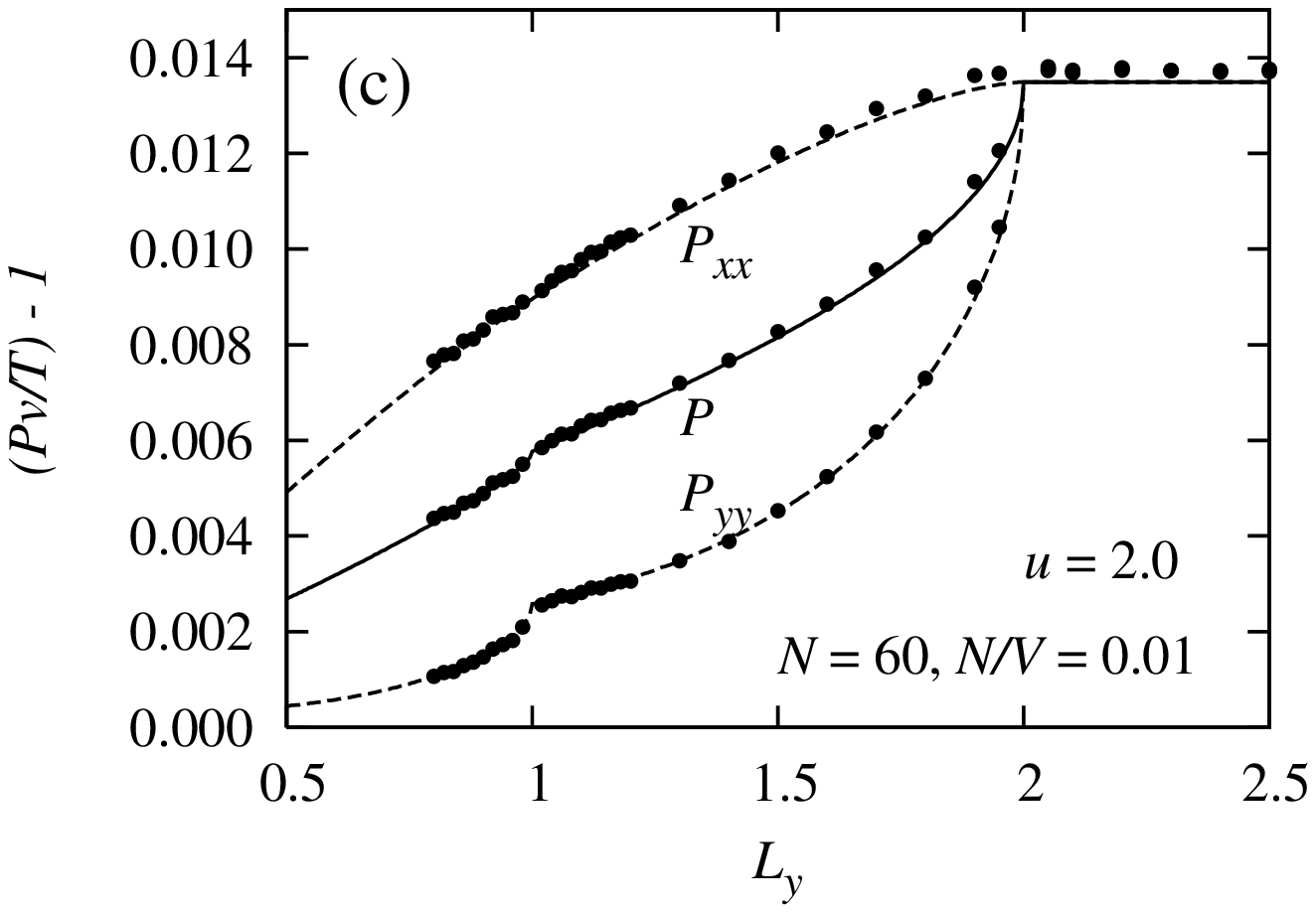}}
 \vspace{-8mm}
\caption{
Comparison of the channel-width dependence of the theoretical
pressures (lines) with numerical simulations (points) for a system  of 60
disks interacting with the step potential of Eq. (\ref{potential1}). The  density
is kept constant, $N/V = 0.01$.
The potential varies from $u=0.5$ (top) to $u=2$ (bottom).
The singularities appear at $L_y = 2$ and $L_y = 1$ as explained
in the main text. Reduced units are used, for which $d$ and the total
energy per particle, $E/N$,  are unity.
 } \label{Fig2}
\end{figure}
In Fig.  \ref{Fig2} the theoretical expressions for the singularity
at $L_y = 2d = 2$ are compared to
computer simulation results for various potential step sizes $u$ as
indicated by the labels. Reduced units are used for which the
particle diameter $d$ and the total energy per particle, $E/N$, are
unity.  The energy $E/N$ is almost exclusively kinetic in nature
with a time-averaged kinetic temperature $T=0.986$ for $u=0.5$
(top), $T=0.994$ for $u=1.0$ (middle), and $T=0.996$ for $u=2$
(bottom figure). These temperatures vary slightly, but
insignificantly, with the channel width $L_y$. The density is kept
constant,  $N/V = 0.01$. As for the case of hard spheres
($u=\infty$) treated already in Ref. \cite{FMP04}, the agreement
between theory and simulation results is very satisfactory.

\begin{figure}
\centering {\includegraphics[width=10cm]{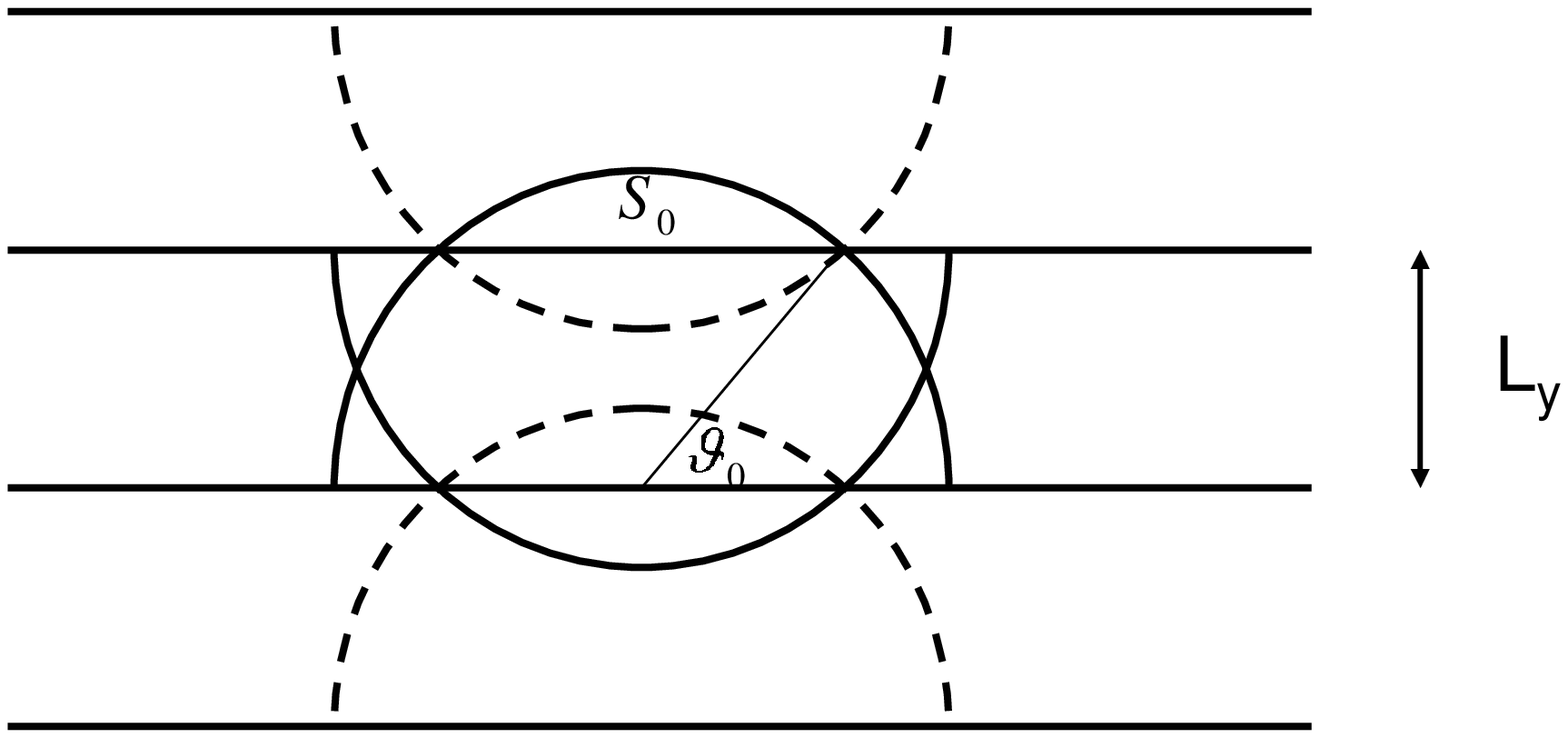}}
\caption{The interaction disks arrangement for $L_y < d$ with periodic boundary
conditions. The overlap area of the disk and its next nearest
neighbor in the $y$ direction is $S_0$.
 } \label{Fig.4}
\end{figure}
As the width of the channel further decreases we expect to have more
singularities to take place at $L_y=d,~2d/3,~d/2,\dots$, which result
from the overlap of an interaction disk with those of further
neighboring disks in the $y$ direction. Let us analyze, for example,
the second singularity in the pressure curve, which takes place at
$L_y=d$. For $d/2 < L_y < d$ the disk of a particle has some overlap
with its nearest and next nearest neighbors images which result from
the periodic boundary conditions in the $y$ direction. In order to
evaluate the overlap integral (\ref{q}) we note that (see Fig. \ref{Fig.4})
\begin{equation}
L_y=d \sin\vartheta_0\,. \label{L_y<1}
\end{equation}
The overlap area between an interaction disk of a particle and that
of its next nearest neighbor in the $y$ direction, $S_0$, is given
by
\begin{equation}
S_0=d^2(\pi -2 \vartheta_0 - \sin{2 \vartheta_0})\,.
\end{equation}
The overlap integral is thus expressed as
\begin{equation}
q(L_y)=(\pi d^2-2S + S_0)(e^{-\beta u}-1) +(S-2S_0)(e^{-2\beta u}-1)
+S_0(e^{-3\beta u}-1) \;.
\end{equation}
Using this expression for the overlap integral, the pressure
components can readily be calculated to yield
%
%
\begin{eqnarray}
\frac {P_{xx}v}{kT} &=& 1 - \frac{q{(L_y)}}{2v} \; \label{pxxnarrow} \\
\frac {P_{yy}v}{kT} &=& 1 -
\frac{q{(L_y)}}{2v}-\frac{L_y}{2v}\sqrt{4d^2-L_y^2}\;(1-e^{-\beta
u})^2-\frac{2 L_y}{v}\sqrt{d^2-L_y^2}\; e^{-\beta u} (1-e^{-\beta u})^2
\; \label{pyynarrow} \\
P  & = & \frac{1}{2} ( P_{xx}  + P_{yy}) \;.
\label{pnarrow}
\end{eqnarray}
As is also shown in Figure \ref{Fig2}, these expressions for the  singularity at $L_y = d = 1$
compare very well with simulation results.
As before, reduced units are used
for which $d$ and $E/N$ are unity. Note that as long as $u>0$ the
coefficient of the singular term $\sqrt{d^2-L_y^2}~$ is positive,
resulting in a positive compressibility just below $L_y=d$.

\subsection{Two-step potential}

In order to analyze the case of disks with an attractive potential, one
has to add a repulsive interaction at short distances to prevent the
collapse of the system into a macroscopic cluster. We thus consider
in this section a two-step potential
\begin{equation}
u(r)= \left\{ \begin{array}{ll} u_1 & \qquad r \leq d \;,\\
u_2 & \qquad d \leq r<D \;, \\ 0 & \qquad r \geq D \;.
\end{array} \right. \end{equation}
where $u_1>0$ represents a repulsive interaction, $u_2$ could be
either positive or negative, and $D>d$ is the outer radius of  $u_2$.
To evaluate the pressure we associate with each
particle two concentric interaction disks, one with radius $d$ and
the other with radius $D$. Two particles only interact with each
other, if the center of the second particle lies within the
interaction disks of the first. It is easy to see that the degree of
overlap of the disks of a particle and those of its nearest neighbor
image resulting from the periodic boundary condition in the y
direction are singular at $L_y=2D, \; d+D$, and $2d$. Thus, the pressure
curve is expected to be singular at these three values of the
channel's width.

We now analyze the pressure curve in more detail and consider
first the upper singularity at $L_y=2D$. For $L_y
> 2D$ the disks of a particle and those of its image do not
overlap. Thus the integral (\ref{q}) yields
\begin{equation}
q(L_y)=\pi (D^2-d^2)(e^{-\beta u_2}-1) + \pi d^2 (e^{-\beta u_1}-1) \qquad
\mbox{for} \qquad L_y>2D \,, \label{q>2D}
\end{equation}
and $q$ is independent of $L_y$. As in the case of a single step
potential, one finds that to leading order in $1/v$ the pressure
tensor is isotropic, $P_{xx}=P_{yy}=P$, with
\begin{equation}
\frac{Pv}{kT}=1 - \frac{q}{2v}\;.
\label{rgt2d}
\end{equation}
For $d+D \leq L_y \leq 2D$, however,  the outer disks
of a particle and its periodic image overlap. As in the case of the single
step potential, the overlap area $S$ is given by
\begin{equation}
S=D^2(\pi -2 \vartheta - \sin{2 \vartheta})\,,
\end{equation}
where $\vartheta$ satisfies (see Fig.~(\ref{Fig1}))
\begin{equation}
L_y=2D\sin\vartheta\,. \label{L_y}
\end{equation}
The resulting overlap integral (\ref{q})~ for $d+D < L_y < 2D$ is
given by
\begin{equation}
q(L_y)=(\pi D^2-\pi d^2-2S)(e^{-\beta u_2}-1) + \pi d^2 (e^{-\beta
u_1}-1) + S(e^{-2\beta u_2}-1) \,.
\label{1+D<q<2D}
\end{equation}
The pressure tensor in this regime is thus found to be
\begin{eqnarray}
\frac {P_{xx}v}{kT} &=& 1 - \frac{q{(L_y)}}{2v} \,,  \label{swpxx}\\
\frac {P_{yy}v}{kT} &=& 1 -
\frac{q{(L_y)}}{2v}-\frac{L_y}{2v}\sqrt{4D^2-L_y^2}(1-e^{-\beta
u_2})^2 \,, \label{swpyy}\\
\frac {Pv}{kT} &=& 1 -
\frac{q{(L_y)}}{2v}-\frac{L_y}{4v}\sqrt{4D^2-L_y^2}(1-e^{-\beta
u_2})^2 \,, \label{swp}
\end{eqnarray}
where $q(L_y)$ is given by (\ref{1+D<q<2D}). As in the case of a
single step potential, $P_{yy}$ exhibits a square root singularity,
while $P_{xx}$ behaves more smoothly, with a weaker 3/2 power
behavior at the transition. The compressibility below the transition
is negative.

%
\begin{figure}
\centering {\includegraphics[width=10cm]{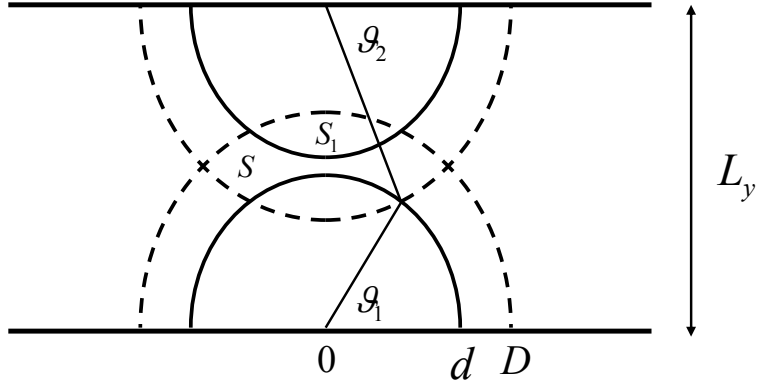}}
\caption{The interaction disks arrangement for the case of a two-step potential
and for $2d \le L_y \le d+D$ with periodic boundary conditions. The
overlap area of the outer disks of two nearest neighbors in the $y$
direction is $S_1$, while the overlap area of the outer disk with
the inner one of its nearest neighbor in the $L_y$ direction is $S_1$.
The angles $\vartheta_1$ and $\vartheta_2$ are indicated .} \label{Fig.5}
\end{figure}
%

Finally, we consider the regime $2d \leq L_y \leq d+D$. In this case
the outer disk of a particle partially overlaps not only with the
outer disk of its periodic image but also with the inner one (see
Fig. (\ref{Fig.5})). The overlap area $S_1$ between the outer and
the inner disks is given by
\begin{equation}
S_1=\frac{1}{2}d^2(\pi-2\vartheta_1)-d^2\sin\vartheta_1\cos\vartheta_1+\frac{1}{2}
D^2(\pi-2\vartheta_2)-D^2\sin\vartheta_2\cos\vartheta_2
\end{equation}
where $\vartheta_1$ and $\vartheta_2$ satisfy (see Fig. (\ref{Fig.5}))
\begin{equation}
L_y=d \sin\vartheta_1 + D\sin\vartheta_2 \,,\qquad \mbox{and} \qquad
d\cos\vartheta_1=D\cos\vartheta_2\,, \label{sumrule}
\end{equation}
and, hence,
\begin{equation}
\sin\vartheta_1 = \frac{1}{2dL_y} (L_y^2 -D^2 + d^2) , \qquad
\sin\vartheta_2 = \frac{1}{2DL_y}(L_y^2 +D^2 - d^2)\,. \label{sin12}
\end{equation}
It is straightforward to express the overlap integral (\ref{q}) in
terms of the overlap areas $S$ and $S_1$ as
\begin{eqnarray}
q(L_y)&=&(\pi D^2-\pi d^2-2S+2S_1)(e^{-\beta u_2}-1) + (\pi
d^2-2S_1)(e^{-\beta u_1}-1)\nonumber \\ &+& (S-2S_1)(e^{-2\beta
u_2}-1) +2S_1(e^{-\beta(u_1+u_2)}-1) \qquad \mbox{for} \qquad 2d\leq
L_y \leq d+D \,.\label{2<q<1+D}\,
\end{eqnarray}
According to Eqs. (\ref{pxxgeneral})  and (\ref{pyygeneral}),  the
pressure components may be expressed as
\begin{eqnarray}
\frac {P_{xx}v}{kT} &=& 1 - \frac{q{(L_y)}}{2v} \,,\\
\frac {P_{yy}v}{kT} &=& 1 -
\frac{q(L_y)}{2v}+\frac{L_y}{2v}(1-e^{-\beta u_2})^2 \frac{dS}{dL_y}
   + \frac{L_y}{v} (e^{-\beta u_2} - e^{-\beta u_1})(1 - e^{-\beta u_2})
 \frac{dS_1}{dL_y} \,,
\end{eqnarray}
where
\begin{equation}
\frac{dS_1}{dL_y} = - 2d\cos(\vartheta_1)=-2D\cos(\vartheta_2)\,.
\label{dS1dL}
\end{equation}
Using Eq. (\ref{sumrule}) and (\ref{sin12}), the channel-width
dependence of the pressure tensor components for $2d < L_y < d+D$ is
finally obtained,
%
%
%
%
%
%
\begin{eqnarray}
\frac {P_{xx}v}{kT} &=& 1 - \frac{q{(L_y)}}{2v} \,, \label{swpxxfin}\\
\frac {P_{yy}v}{kT} &=& 1 -
\frac{q(L_y)}{2v}-\frac{L_y}{2v}\sqrt{4D^2-L_y^2}(1-e^{-\beta
u_2})^2
 -\frac{1}{v} (e^{-\beta u_2} - e^{-\beta u_1})(1 - e^{-\beta u_2})
   \nonumber \\ &&\times
(4L_y^2D^2+4L_y^2d^2+4D^2d^2-L_y^4-D^4-d^4)^{1/2} \,,
\label{swpyyfin}
\end{eqnarray}
where $q(L_y)$ is given by (\ref{2<q<1+D}).

In order to establish the nature of the singularity at the
transition point   $ d + D$, we expand $dS_1/dL_y$ in Eq.
(\ref{dS1dL})  in terms of the small dimensionless offset  $\epsilon = (d + D -
L_y)/d$. Introducing the small angles
\begin{equation}
\vartheta_1=\frac{\pi}{2} - \delta\vartheta_1 \,,\qquad \vartheta_2=\frac{\pi}{2}
-\delta\vartheta_2\,,
\end{equation}
which, according to Eq. (\ref{sumrule}), are related to $\epsilon$ by
\begin{equation}
(\delta\vartheta_1)^2=\frac{2D}{(d+D)} \epsilon \,,\qquad
(\delta\vartheta_2)^2=\frac{2d^2}{D(d+D)} \epsilon\,,
\end{equation}
we finally obtain
\begin{equation}
\frac{dS_1}{dL_y} = - d \sqrt{ \frac{8 D}{d+D}}  \sqrt{\epsilon}
\end{equation}
It is readily seen that $dS_1/dL_y$ and, hence, $P_{yy}$, exhibit a
square root singularity. As in the case of a single step potential,
the singularity in $P_{xx}$ originates from $S_1$ rather than from
its derivative. Hence $P_{xx}$ exhibits a weaker $3/2$ singularity
at $L_y=D+d$. It is interesting to note that, depending on the
values of the interaction parameters $u_1$ and $u_2$, the
compressibility just below the transition could be either positive
or negative.

\begin{figure}
\centering
{\includegraphics[width=10cm]{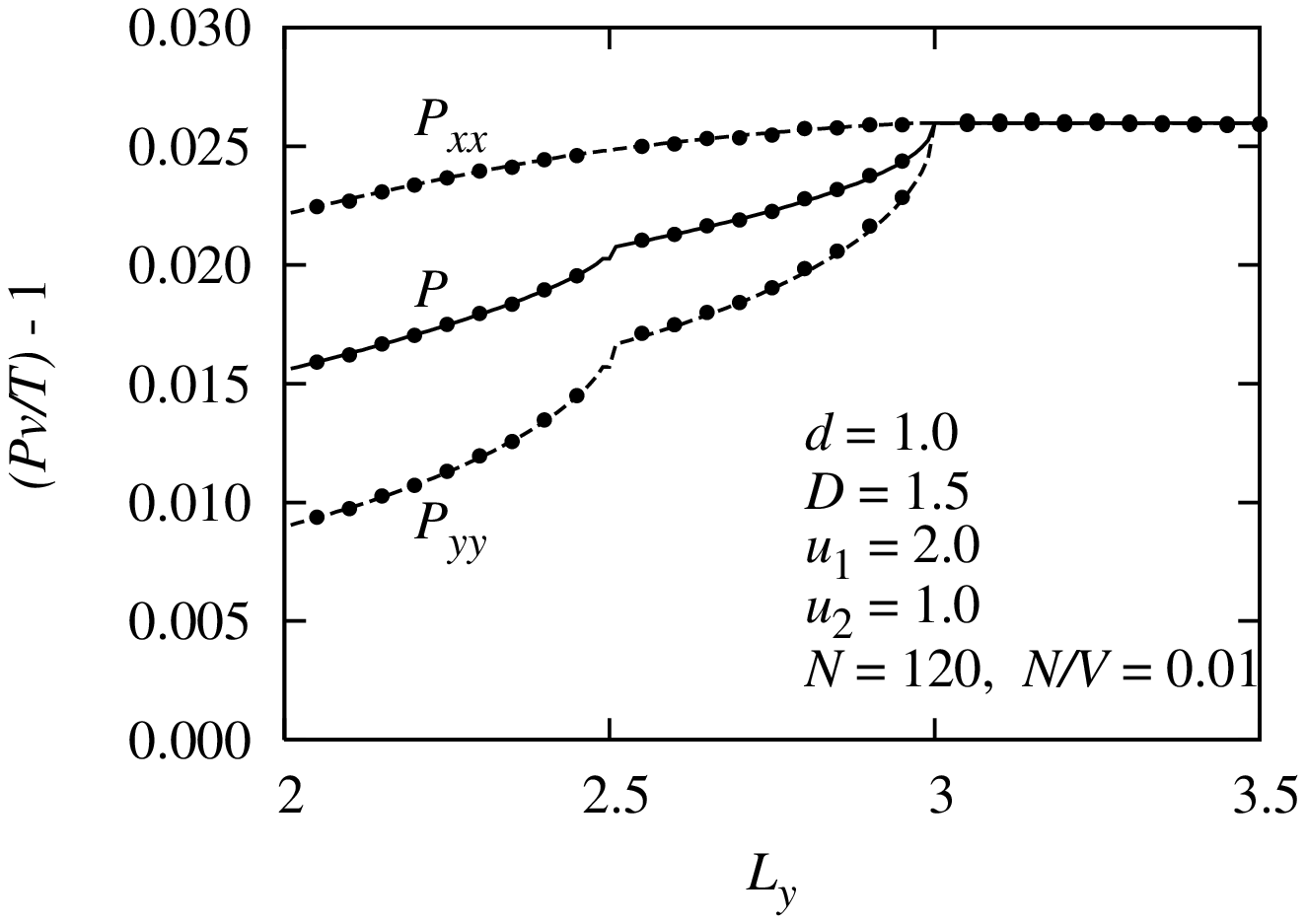}\\
\vspace{-3mm}
 \includegraphics[width=10cm]{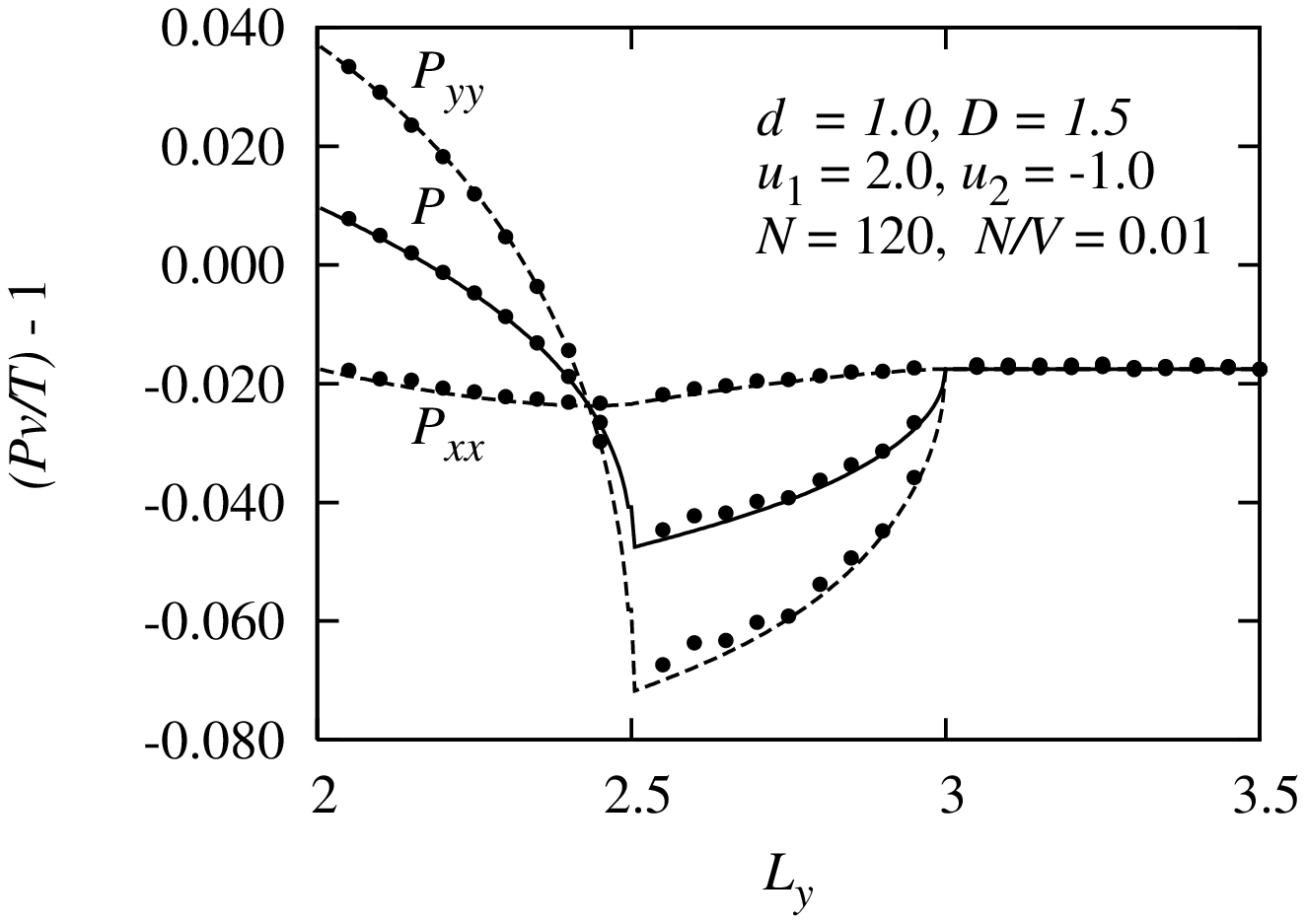}}\\
 \vspace{-3mm}
\caption{Channel-width dependence of the pressures for the two-step
potential case with the following set of parameters:  $d = 1$, $D =
1.5$, $N = 60$, $N/V = 0.01$ and $E/N=1$. The top panel corresponds
to a potential with two positive steps, $u_1 = 2$ and $u_2 = 1$,
whereas the lower panel is for an attractive outer shell with $u_1 =
2$ and $u_2 = - 1$.
 } \label{Fig6}
\end{figure}
In Fig. \ref{Fig6} we compare the respective theoretical expressions
-- Eq. (\ref{rgt2d}) for $L_y > 2D$, Eqs. (\ref{swpxx} - \ref{swp})
for $d+D < L_y < 2D$, and Eqs. (\ref{swpxxfin}, \ref{swpyyfin}) for
$2d < L_y < d+D$ --
to computer simulation results (dots) for $N=60$ particles in a
narrow channel of width $L_y$ with periodic boundaries both in $x$
and $y$ directions. We use reduced units for which $E/N$ and $d$ are
unity. The outer diameter D = 1.5. In the top panel, results for a
two-step potential with $u_1 = 2$ and $u_2 = 1$ are shown. The lower
panel corresponds to a true square well potential with $u_1 = 2$ and
$u_2 = - 1$. The agreement is very good in all cases.

A similar analysis may be carried out near the third singularity
which takes place at $L_y=2d$.


\subsection{Power-law potential}
\label{PLP}

\begin{figure}
\centering
{ \includegraphics[width=10cm]{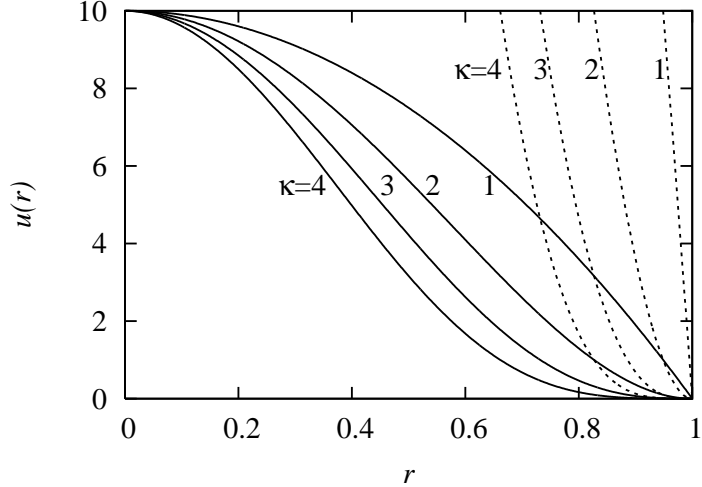}}
 \vspace{-3mm}
\caption{Potentials for $A = 10$ (smooth curves) and $A=100$ (dashed
curves)  for various  values of $\kappa$ as indicated by the labels.
 } \label{pot}
\end{figure}
%
We now consider a soft potential which vanishes (continuously) at
the disk boundary,
\begin{equation}
u(r)= \left\{ \begin{array}{ll} A\left(1-\left(\frac{r}{d} \right)^2 \right)^\kappa & \qquad r \leq d \\
0 & \qquad r>d
\end{array} \right.
\label{contin}
\end{equation}
where the two parameters $A > 0$ and $\kappa > 0$ are constants. In
Fig. \ref{pot} we show a few of such potentials for $A = 10 E_0$
(smooth lines) and $A = 100 E_0$ (dashed lines) for various $\kappa$
as indicated by the labels. Here $E_0 \equiv E/N = \left[ \sum_i
p_i^2/ 2m + \sum_i \sum_{j>i} u(r_{ij})\right] /N $ is the {\em
total} energy per particle. For our numerical work we use reduced
units, for which the diameter $d$, the particle mass $m$, and $E_0$
are unity.

Let us analyze the nature of the singularity  of the pressure
components slightly below $2 d$. The singularity arises from
integrating the Mayer  function $f$ in the overlap region of the two
disks. For $\epsilon \equiv (2d -L_y)/d  \ll 1$, the function $f$ is
small in this region and may thus be expanded in powers of $u(r)$.
To second order in $u$, $f(r)\simeq -\beta u(r) + (1/2) (\beta
u(r))^2$. The singularity in the integral (\ref{q}) arises from the
non-linear term in $f$, which is of the order $\epsilon^{2\kappa}$
in the overlap region. Since according to (\ref{area}) the area of
this region scales as $\epsilon^{3/2}$ for small $\epsilon$, the
singular contribution to the integral (\ref{q}), and hence to
$P_{xx}$, scales as $\epsilon^{2\kappa+3/2}$. On the other hand, the
pressure $P$ and its $P_{yy}$ component scale as
$\epsilon^{2\kappa+1/2}$. Thus, for small enough $\epsilon$ we
expect
\begin{equation}
P  = c_1 \left[ 1 - c_2 (2d - L_y)^{2 \kappa+1/2} \right] \; ,
\label{pfit}
\end{equation}
and similarly for $P_{yy}$, where $c_1, c_2$ are constants. In the
scaling form for $P_{xx}$, the exponent is $2\kappa +3/2$, and the
singularity is weaker.

To test this scaling form, we carried out numerical
simulations of the model.
 In selecting the parameters $A$ and $\kappa$ most appropriate for
numerical simulations, one should take into account two competing
trends. On the one hand, the singular part of the pressure is
expected to be more pronounced for large amplitude $A$ and small
exponent $\kappa$. On the other hand, as we argue below, the
channel-width interval, where the scaling form (\ref{pfit}) is expected to
hold, is larger for small $A$ and large $\kappa$. Thus, in order to
observe the scaling behavior one has to choose intermediate values
of these two parameters.

To estimate the scaling interval $L_{y,\min} < L_y < 2d$ over which
the scaling form (\ref{pfit}) is expected to hold, we note that
during a typical collision two particles penetrate each other up to
a depth $\delta = d - r_0$, where $r_0$ is estimated from $u(r_0) =
E_0$,  $ r_0 = d \sqrt{ 1 - (E_0/A)^{1/\kappa}} $. The expansion of
the Mayer function  to second order in $u$ fails, if the third-order
term starts to contribute more than, say, 10 \%. This failure only
happens for particle separations  smaller than $r_0$, the typical
separation at maximum penetration, and, hence, for untypical
high-energetic collisions.  For typical energies and penetrations,
particles will pass each other in the channel and contribute to the
pressure scaling, if the thermally possible penetration depth
$\delta$ exceeds the interaction-disk overlap $d \epsilon$ due to
the periodic boundaries. The upper bound for $\epsilon$ is thus
estimated to be $\epsilon_{\max} = \delta/d  = (d- r_0)/d$, and the
minimum channel width for which scaling is expected to hold becomes
$L_{y,\min} = 2d -  d\epsilon_{\max} = d + r_0$. Thus, the scaling
interval decreases   with $A$ and increases with $\kappa$. We find
that $\kappa =2$ and $A= 10$ are a suitable choice, which gives a
reasonable scaling range, and we consider this case first. Note that
the choice $\kappa > 1$  also offers the slight numerical advantage
that the particle force is continuous and vanishes at $r = d$.

The simulation results for the pressures  with  potential parameters
$ A = 10 E_0$ and $ \kappa = 2$ are shown by the dots in the bottom
panel of    Fig. \ref{pressure_e400}. In the simulation we used 20
particles at a density $N/V = 0.01$.
\begin{figure}
\centering
{\includegraphics[width=10cm]{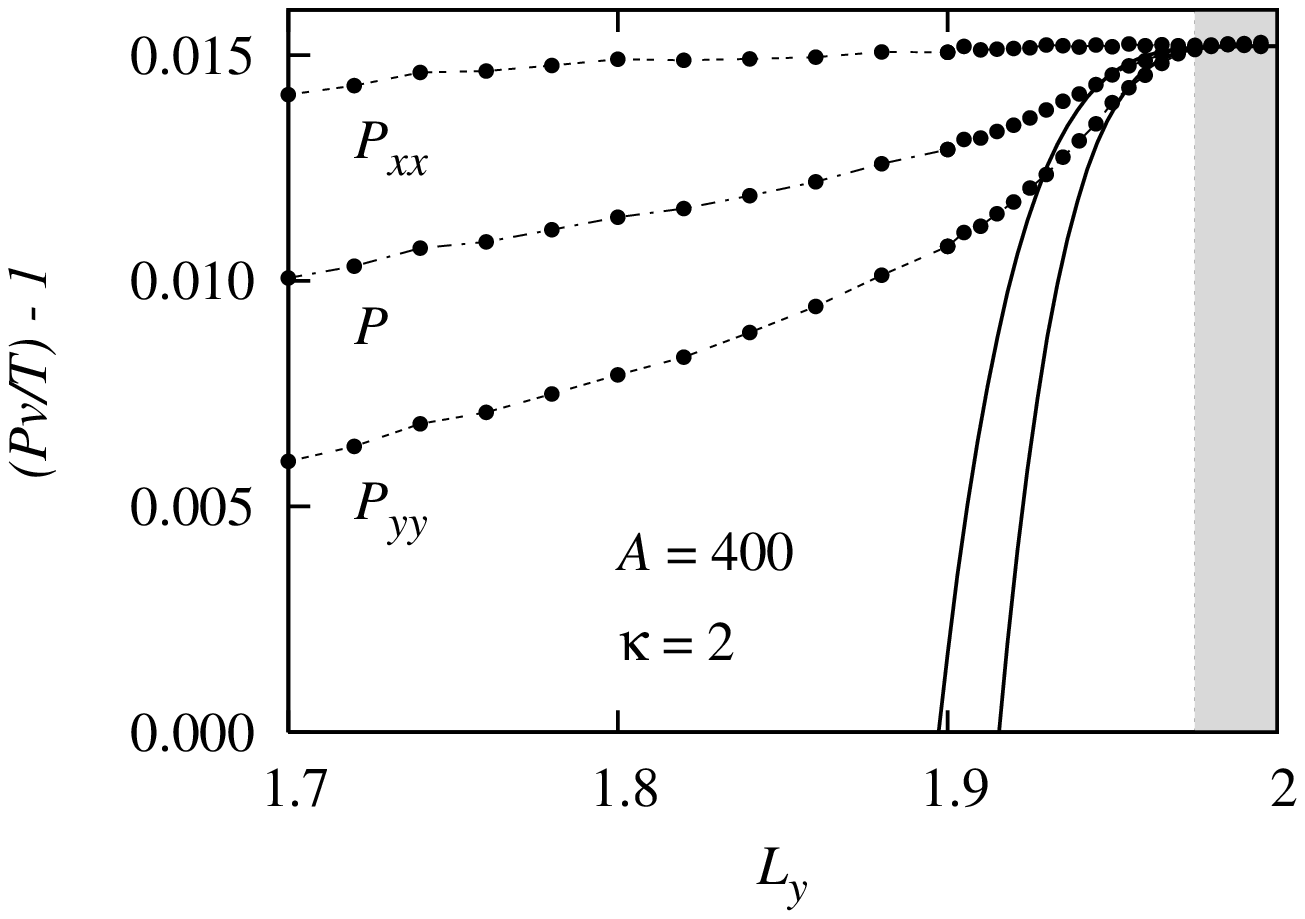}}\\
\vspace{-6mm}
{\includegraphics[width=10cm]{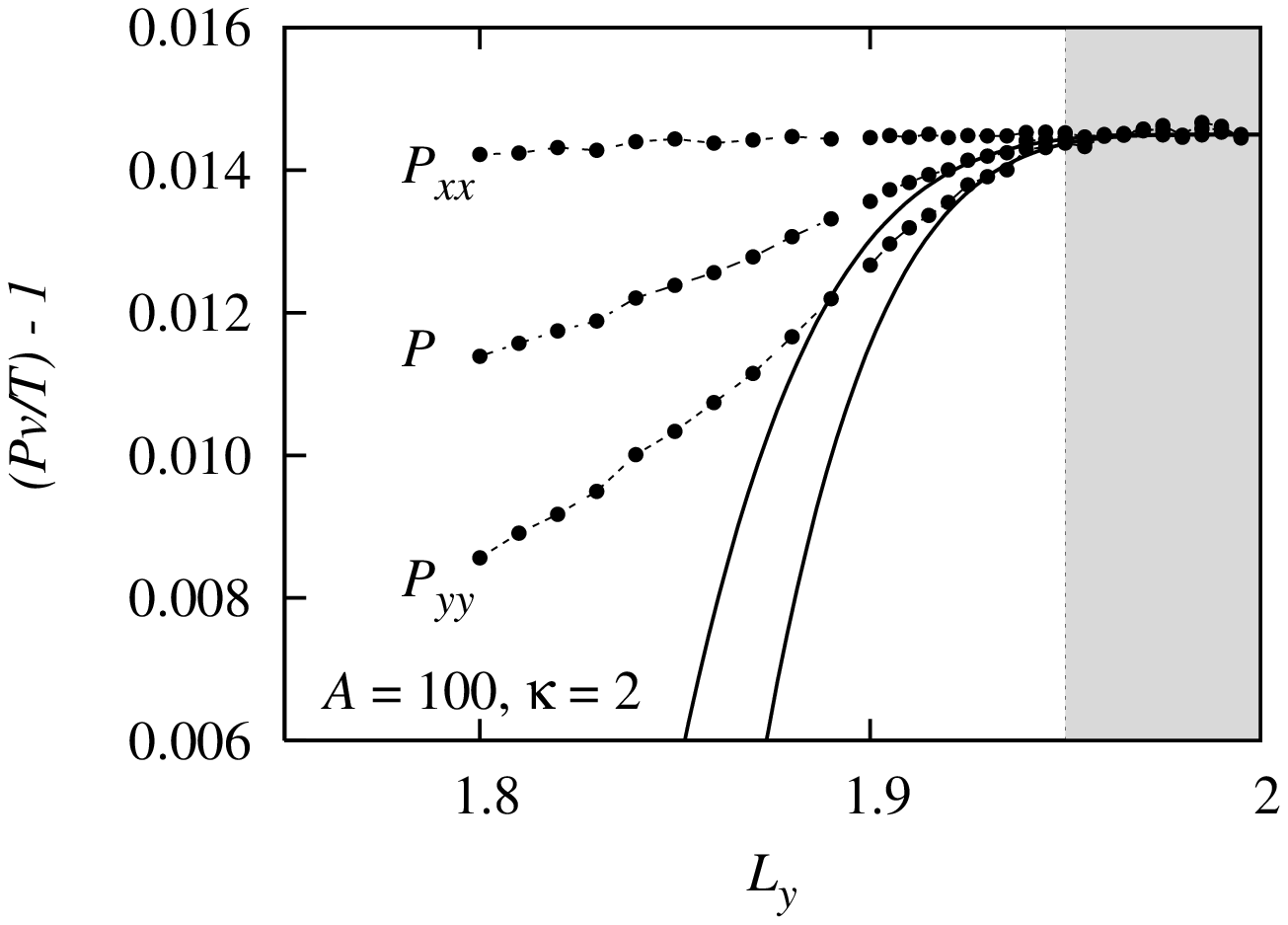}}\\
 \vspace{-6mm}
{\includegraphics[width=10cm]{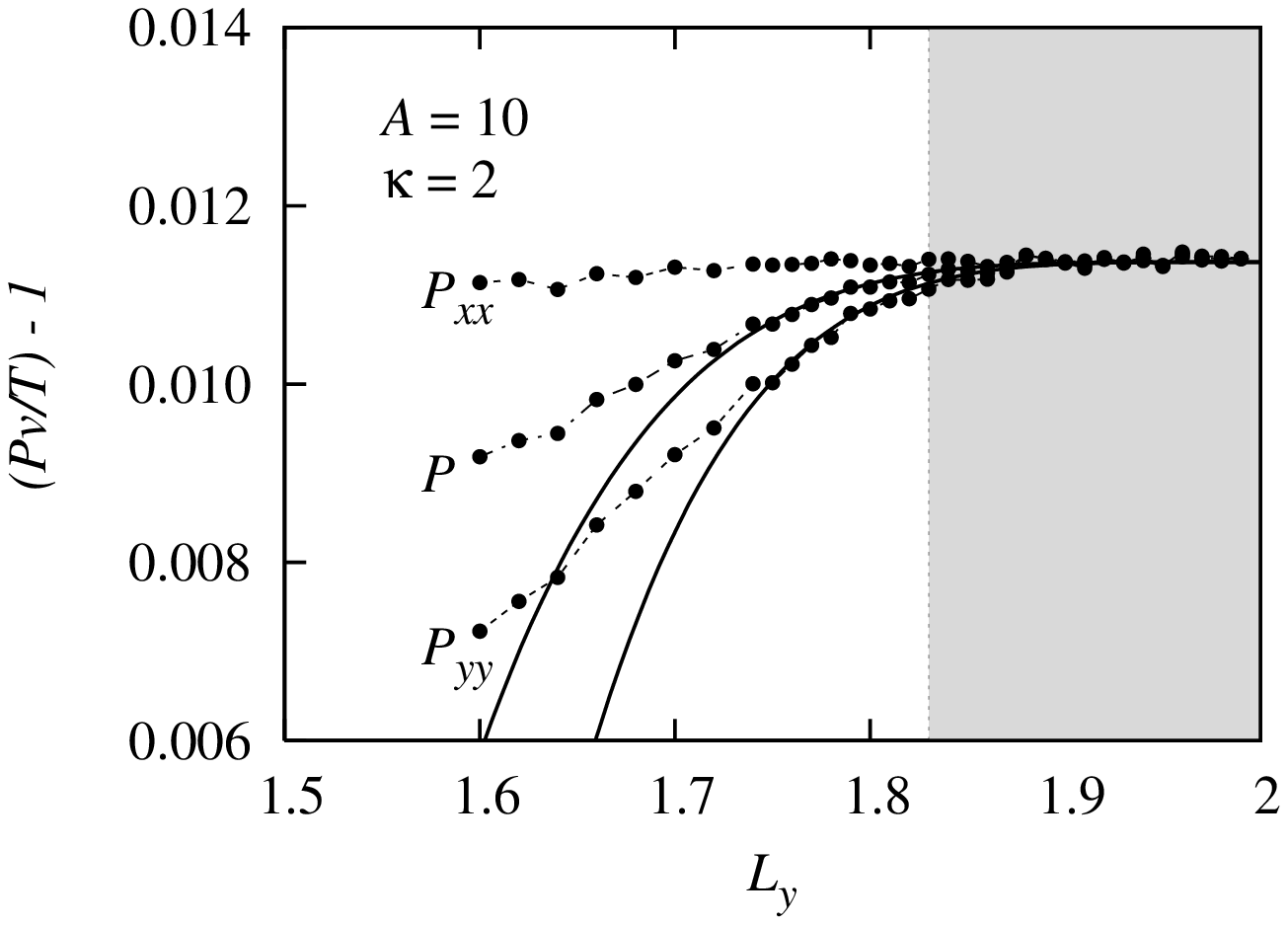}}
 \vspace{-8mm}
\caption{Simulation results for the pressures as a function of the
channel width $L_y$ with periodic boundaries. The potential
parameter $A$ varies from $A = 400$ (top) to $A = 100$ (middle) and
$A=10$ (bottom),  and $\kappa = 2$. In this figure $N/V = 0.01$, and
$N = 20$. The shaded areas indicate (very conservative) estimates of
the scaling regimes. The smooth lines are a fit of Eq. (\ref{pfit})
for $P$ and $P_{yy}$ to the data in the shaded regime. Reduced units
are used for which $d$ and $E_0$  are unity.} \label{pressure_e400}
\end{figure}
The estimated scaling interval, $L_{y,\min} \approx 1.83 d <  L_y  <
2d $, is indicated by the shaded area. The smooth lines are a fit of
Eq. (\ref{pfit}) to the numerical data points for  $P$ and $P_{yy}$
in that range. It shows that our estimate is  rather conservative,
since the fits represent the data points reasonably well in a
slightly wider interval $1.76 d \le L_y \le 2 d$. The kinetic energy
per particle is about $0.998 E_0$ and varies only marginally with
$L_y$. The scaling is more convincingly demonstrated in Fig.
\ref{scalcon}, where
\begin{figure}
\centering
{ \includegraphics[width=10cm]{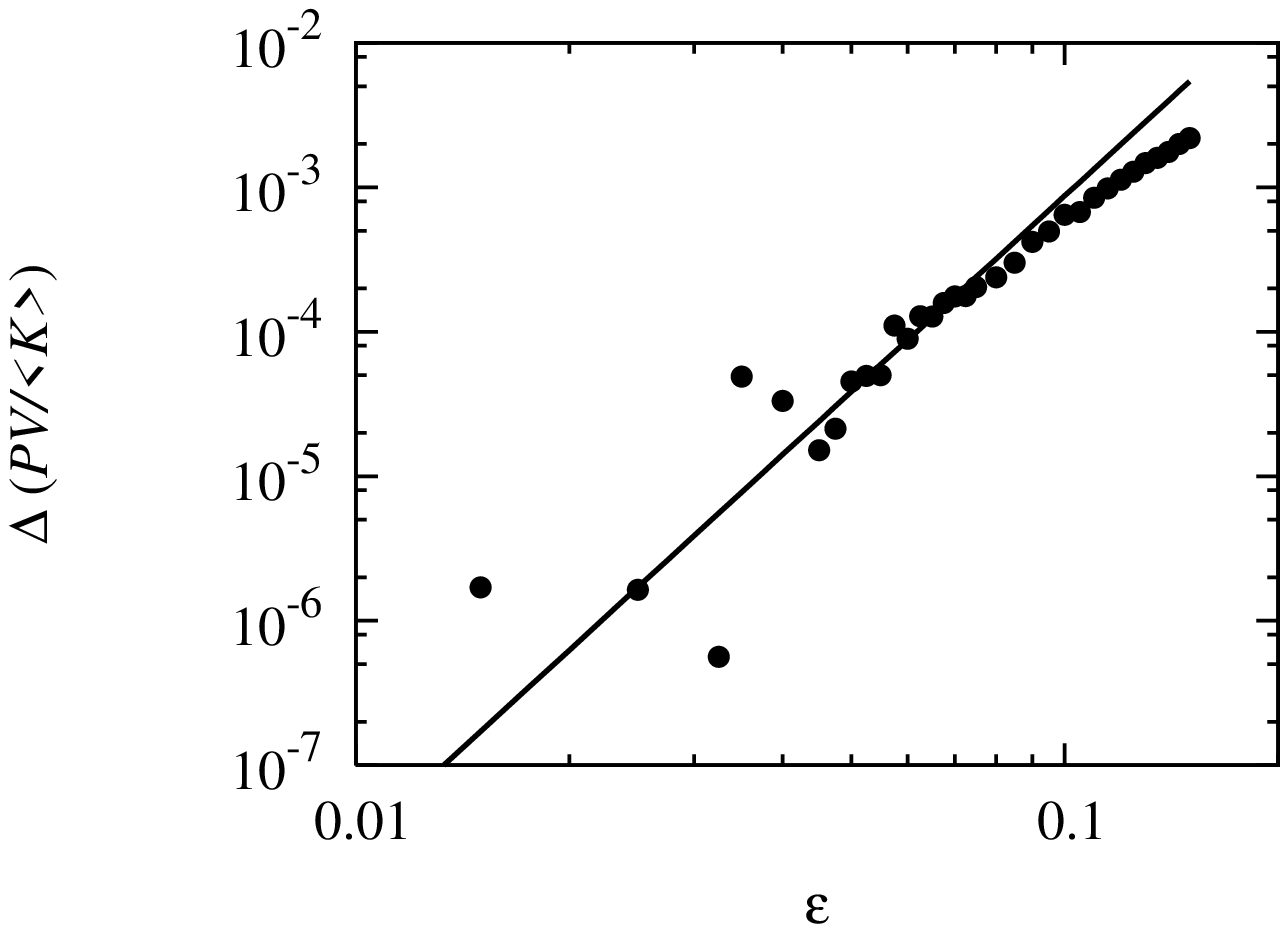}}
 \vspace{-3mm}
\caption{Scaling of the transverse pressure $P_{yy}$ for the power-law potential model
with $A = 10$ and $\kappa = 2$ below the critical channel width $L_y = 2$.
Here, $\epsilon = (2 - L_y)$. The slope of the straight line corresponds to the
theoretically expected scaling, $\kappa + (1/2) = 4.5$.
Reduced units are used as explained in the main text.
 } \label{scalcon}
\end{figure}
the $\epsilon$-dependence of the singular part,
$\Delta \frac{PV}{\langle K \rangle} \equiv  \left[ \frac{PV}{\langle K \rangle} \right] -
     \left[ \frac{PV}{\langle K \rangle} \right]_{\epsilon = 0} $, for $P_{yy}$
is shown. The straight line indicates the expected scaling with  the
power 4.5.  For $\epsilon > 0.07$ corresponding to $L_y < 1.86 \sigma$,  the scaling breaks
down as expected.

 Next, we consider a potential with $A=100 E_0, \; \kappa = 2$, which is  shown in
 Fig. \ref{pot}, and which resembles more realistic repulsive potentials.
 The channel-width dependence of the pressures is shown in the middle panel
 of   Fig. \ref{pressure_e400}. The estimated scaling range is much narrower
 than before, $L_{y,\min}  = 1.95 d$, and is indicated by the shaded area.
The smooth lines are fits of Eq.   (\ref{pfit}) to the data points for $P$ and $P_{yy}$
in that interval and confirm  that our
scaling-range estimate is rather conservative. The kinetic energy per particle
is around $0.999 E_0$ in this case, and varies marginally with $L_y$.

Finally, we consider the limiting case of a rather steep potential such as
for $A = 400 E_0$ and $ \kappa = 2$ (much steeper than the $\kappa = 2$ curve
in Fig. \ref{pot}), which
already resembles that of hard disks and, therefore, should give
a pressure variation with  $L_y$  similar to that found in Ref. \cite{FMP04}.
The results are shown in the top panel of Fig. \ref{pressure_e400}.
The average kinetic energy per particle is $0.9997 E_0$.
For $L_y  < 1.95 d$ the   pressure curves are indeed very similar to those of a
hard-disk gas of 20 disks at the same density and at unit kinetic energy per particle
as is shown in Fig. 3 of Ref. \cite{FMP04}. Differences  appear for channel widths
very close to $2 d$, which are due to the expected scaling.
The estimated scaling range  is very narrow, $L_{y,\min}  = 1.975 d$ as
indicated by the shaded area in the top panel of Fig. \ref{pressure_e400}.
But a fit of Eq. (\ref{pfit}) in the range $1.96 d\le L_y \le 2 d$ represents the
data points for $P$ and $P_{yy}$ reasonably well in that range as is
shown by the smooth lines.

Before closing this section, we provide some details about the molecular
dynamics simulations. They were carried out with a hybrid code
 combining the advantages of the event-driven algorithm for hard
 particles during the forceless streaming stage with the simplicity of
 a time-stepping integration scheme during the collision of
 two or more particles. The beginning of each pair collision was determined
 as in the event-driven algorithms of the previous sections. During  the
 collisions the equations of motion were integrated with a fourth order Runge-Kutta
 scheme. The end of each pair collision was determined by
 interpolation with a spatial uncertainty of less than 10$^{-8}$ reduced units.
 The moment the last interacting particles separate,
 another streaming move is initiated. This method is particularly suited
 for low densities. It even allows to accurately follow the trajectory for
 models with discontinuous forces. Periodic boundaries are used. In
 most cases a trajectory was followed for two million reduced time units
 $\sqrt{md^2/E_0}$.

%
\section{Channels with reflecting boundary conditions}
\label{reflecting-boundaries}
%

\subsection{Soft disks: single step potential}

In this section we calculate the pressure components of soft disks
in a narrow rectangular box with elastic reflecting boundary
conditions in the $y$ direction. Since we are interested in the
narrow channel limit where the length of the box is much larger than
the width, the system is not sensitive to the boundary conditions in
the $x$ direction. For simplicity we take periodic boundary
condition in this direction. We consider disks of diameter $d$, which
interact with each other via the square well
potential of Eq. (\ref{potential1}), but with an additional  $\delta$-function
at the center of the particles:
\begin{equation}
u(r)= \left\{ \begin{array}{ll} \delta(r) + u & \qquad r \leq d \;,\\
0 & \qquad r>d \;,
\end{array} \right.
\label{single-step-potential}
\end{equation}
This $\delta$ function does not affect the particle-particle interactions,
but it is responsible for the elastic reflections from the boundary,
which confine the disk centers to the volume $V = L_x L_y$, where
$L_y$ is referred to as the channel width.

As a result of the reflecting boundary conditions, the system is no
longer translationally invariant, and the overlap integral (\ref{q})
corresponding to the second virial coefficient is replaced by
\begin{equation}
q(L_y)=\frac{1}{L_y}\int f_{12} d^2{r_{1}}d^2{r_{2}} ~ .
\label{q_reflecting}
\end{equation}
As was done in the case of periodic boundary conditions, with each
particle we associate an interaction disk with a radius $d$. Two particles interact
with each other only if the center of a particle is within the interaction disk of the
other.

In the case $L_y>d$ the overlap integral is given by (see the top panel
of Fig. \ref{overlap_box2})
\begin{equation}
q(L_y)=\frac{1}{L_y}\left[\pi d^2 L_y - 2\int_0^d S(\vartheta,d)dy
\right](e^{-\beta u}-1)~,
\end{equation}
where $S(\vartheta,d)$ is the segment of a circle of radius $d$
corresponding to a central angle $2\vartheta$,
\begin{equation}
S(\vartheta,d)=d^2h(\vartheta)~, \qquad  h(\vartheta)=\vartheta
-\sin\vartheta \cos\vartheta ~,\qquad \mbox{and} \qquad \cos
\vartheta =\frac{y}{d} ~. \label{S-y}
\end{equation}
Evaluating the integral one obtains
\begin{equation}
q(L_y)=\frac{1}{L_y}\left( \pi d^2 L_y -\frac{4}{3}d^3
\right)(e^{-\beta u}-1)~. \label{q-reflecting>}
\end{equation}
%
%
\begin{figure}
\centering
{\includegraphics[width=8cm]{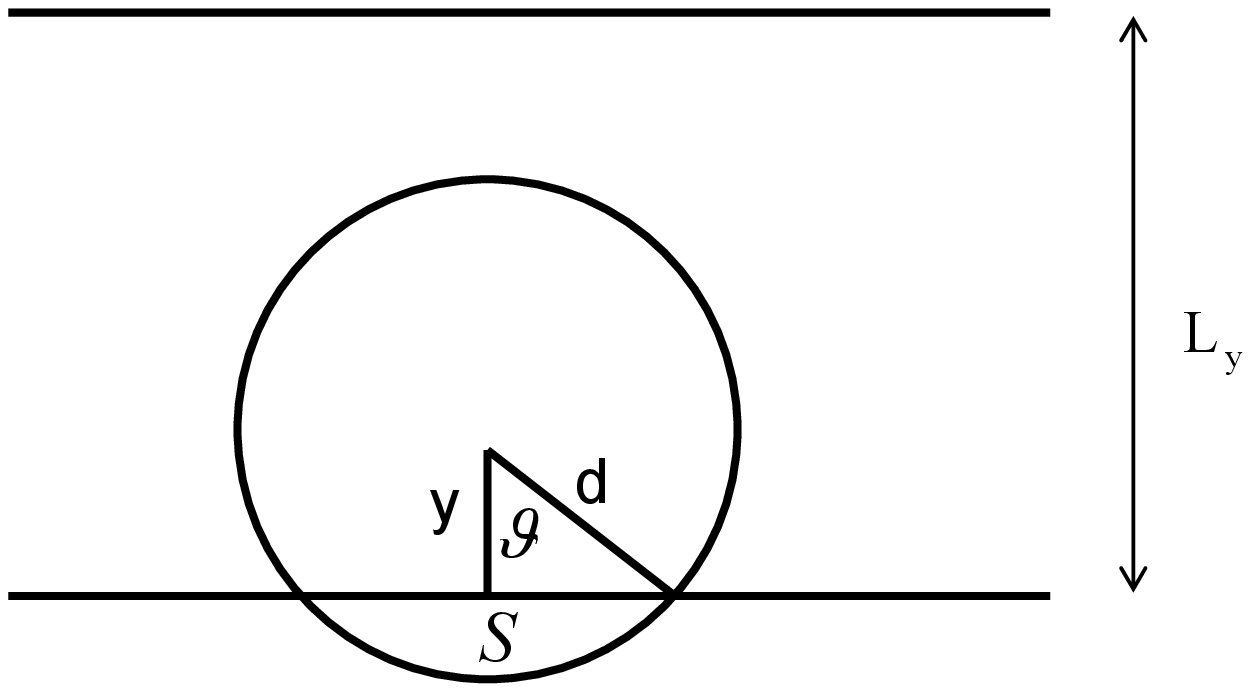}}\\
\vspace{3mm} {\includegraphics[width=8cm]{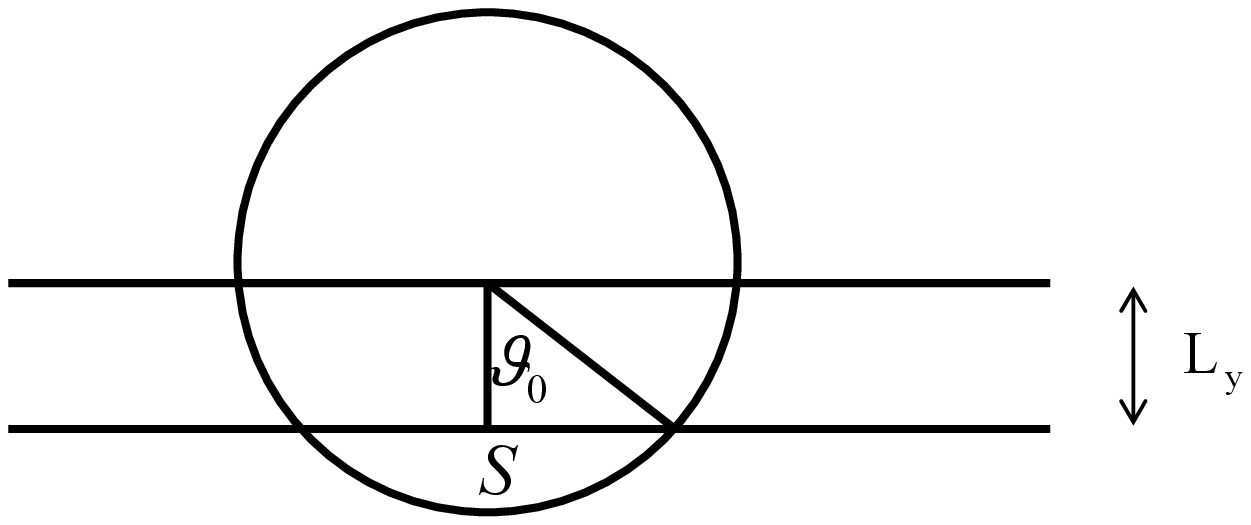}}
\vspace{-3mm} \caption{Interaction disks for the square well
potential in Eq.  (\ref{potential1}) for $L_y > d$ (top), and $L_y <
d$ (bottom) in a channel with reflecting boundary condistions.}
\label{overlap_box2}
\end{figure}

%

For $L_y<d$ the overlap integral is given by (see the bottom panel of Fig.
\ref{overlap_box2})
\begin{equation}
q(L_y)=\frac{1}{L_y}\left[\pi d^2 L_y - 2\int_0^{L_y} S(\vartheta)dy
\right](e^{-\beta u}-1)~,
\end{equation}
where, as in the previous case, $S(\vartheta)$ and $y$ are defined
by Eq. (\ref{S-y}). The integral can be readily evaluated to yield
\begin{equation}
q(L_y)=\frac{1}{L_y}\left[ \pi d^2 L_y
-2d^3\left(\frac{2}{3}+\vartheta_0 \cos \vartheta_0 - \sin \vartheta
+\frac{1}{3} \sin^3 \vartheta_0 \right) \right](e^{-\beta u}-1)~,
\label{q-reflecting<}
\end{equation}
with
\begin{equation}
\cos \vartheta_0 = \frac{L_y}{d} \;.
\end{equation}
The pressure components may be evaluated by using equations
(\ref{q-reflecting>},\ref{q-reflecting<}) in the general expressions
(\ref{pxxgeneral},\ref{pyygeneral}).

This analysis demonstrates that the pressure components exhibit a
singularity at a channel width $L_y=d$. The nature of the
singularity is obtained by expanding the overlap integral in
$1-L_y/d \equiv \epsilon$ for small $\epsilon>0$. In this limit one
has $\vartheta_0 \simeq \sqrt{2\epsilon}$ and $q(L_y)\simeq
-{2/15}\vartheta_0^5$. Thus, for small $\epsilon$ the singular part
of the overlap integral $\delta q(L_y)$ satisfies
\begin{equation}
\delta q(L_y)\simeq \epsilon^{5/2}~.
\end{equation}
As a result, $P_{xx}$ exhibits a $5/2$ singularity at $L_y=d$, and
its third derivative with respect to $L_y$ diverges, as $L_y$
approaches $d$ from below. On the other hand, $P_{yy}$ exhibits a
stronger, $3/2$, singularity, as it is related to the derivative of
the overlap integral with respect to $L_y$.

\begin{figure}
\centering
{ \includegraphics[width=10cm]{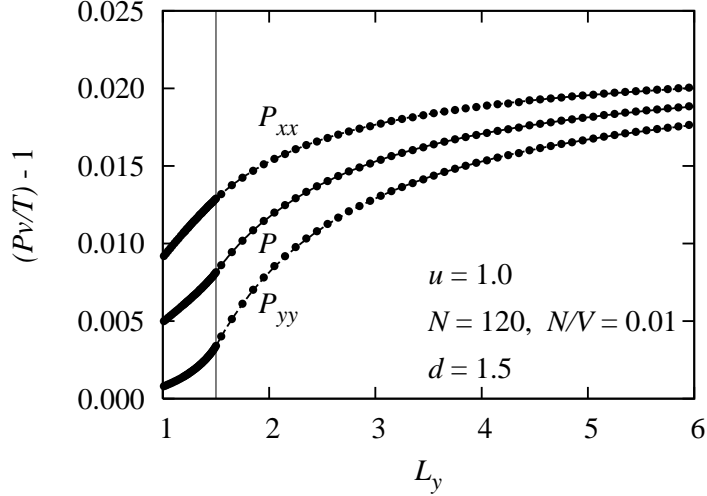}}
 \vspace{-3mm}
\caption{Channel-width dependence of the pressures
for $N=60$ particles, which interact with each other with
the weak repulsive step potential of
Eq. (\ref{potential1}),  where $ u = 1$. Reflecting
boundary conditions are used as described in the main text.
The particle diameter $d = 1.5$ in the reduced units applied, and the energy
per particle, $E/N$, is unity. Keeping the particle density constant,
$N/V = 0.01$,  the temperature varies slightly with $L_y$.  }
\label{Fig8}
\end{figure}
A comparison between theoretical and simulation results is provided
in Fig. \ref{Fig8} for a {\em positive} step potential. We use
reduced units, in which the particle diameter $d$ is 1.5, and for
which the total energy per particle, $E/N \equiv E_0$, is unity. In
these units, we choose for the potential  $u = 1$. In the simulation
we studied $N=120$ particles enclosed in a box with reflecting
boundaries both parallel and perpendicular to the channel axis,
such that the centers of the particles are confined to the volume
$V = L_x L_y$ (and the particle disks to the volume $(L_x + d)(L_y+d)$).
Varying the channel width $L_y$, the density is kept constant, $N/V = 0.01$.
The singular width, $d$, is indicated by the vertical line.  The agreement between the points
from the simulation and the theoretical smooth lines corresponding
to  $P_{xx}, P_{yy}$ and $P = (P_{xx} + P_{yy})/2$ is nearly
perfect.

To demonstrate the scaling directly, we plot in Fig. \ref{scalsteppot}
the computer simulation results for the singular pressure contributions of $P_{xx}$
and $P_{yy}$ below the singular channel width $L_y = d$. To do this,  we note that
the non-singular contribution to the  overlap integral Eq.(\ref{q-reflecting<})  for
$L_y < d$ is given by Eq. (\ref{q-reflecting>}), also  evaluated at $L_y < d$. The
corresponding non-singular  (NS) pressure contributions then follow from
Eqs. (\ref{pxxwide})  and (\ref{pyywide}) with $q(L_y)$ taken from
Eq. (\ref{q-reflecting>}). If this non-singular part is subtracted from the
pressures determined by the simulations, a plot of
$$
Z_{\alpha \alpha}  \equiv  \frac{W_{\alpha \alpha}} {\langle K \rangle} -
\left[ \frac{P_{\alpha \alpha} v}{kT} -1 \right]_{\mbox{NS}} \;  \;, \alpha \in \{x,y\}
 $$
as a function of  the distance from the singularity, $\epsilon =  (d - L_y)/d$,
reveals the expected scaling for the $xx$ and $yy$ pressure components.
This is demonstrated by the
straight lines in the log-log plot  of Fig. \ref{scalsteppot},  which are fully
consistent with the expected scaling, 3/2 for $P_{xx}$, and 5/2 for $P_{yy}$.
\begin{figure}
\centering
{ \includegraphics[width=10cm]{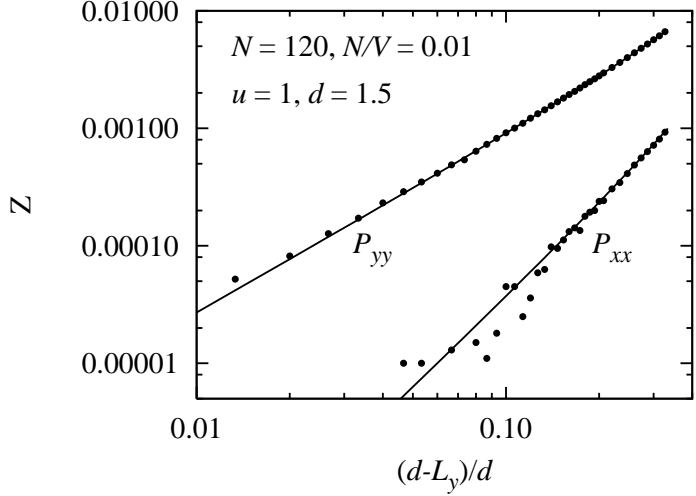}}
 \vspace{-3mm}
\caption{Channel-width scaling of the singular pressure contributions, which are obtained by
subtracting the respective
non-singular contributions from the computer simulation results for the pressures (points).
The straight lines indicate the theoretically expected scaling. For details we refer to the
main text. }
\label{scalsteppot}
\end{figure}

     It is interesting to note that a system with a purely negative box potential, $u < 0$,
is thermally unstable and tends to form clusters. The maximum
entropy state consists of  a single cluster of overlapping disks,
floating in the gas of the remaining particles. As a consequence,
the specific heat is negative \cite{Thirring,Hertel,TNP}, and the
temperature is increased due to the conversion of potential energy
into kinetic. Such a property may only arise in the microcanonical
ensemble and is familiar for gravitational systems. However, the
attracting force need not be of long range \cite{PNT90}. A negative
specific heat may even be observed for quantum-mechanical Coulomb
systems \cite{TNP} and in experiments on nuclear fragmentation
\cite{agostino} and atomic clusters \cite{schmidt,maerk}.
In a preliminary study, we have observed the clustering for $ u = -1$, but we do not
consider this case in more detail, because the times for
reaching thermodynamic equilibrium are excessively long.

%
\subsection{Soft disks with a hard core: two-step potential}
\label{soft_disks_hard_core}
%

In this section we consider particles which interact with a two-step
potential
\begin{equation}
u(r)= \left\{ \begin{array}{ll} \infty & \qquad r \leq d \\
u & \qquad d \leq r<D \\ 0 & \qquad r \geq D
\end{array} \right. \label{2step_potential}\end{equation}
The soft potential $u$ may be either attractive or repulsive. The disks interact
with the walls of the
channel only by the hard core interaction with diameter $d$ (In the previous
section this potential collapses into a $\delta$ function). As before, we
assume reflecting boundary conditions in the $y$ direction  as indicated in
Figs. \ref{overlap_box3} and \ref{overlap_box4}, and
periodic boundary conditions in the $x$ direction. Thus, the volume
accessible to the centers of the particles is given by $V = (L_y - d) L_x$.
\begin{figure}
\vspace{-4cm} \centering {
\includegraphics[width=12cm]{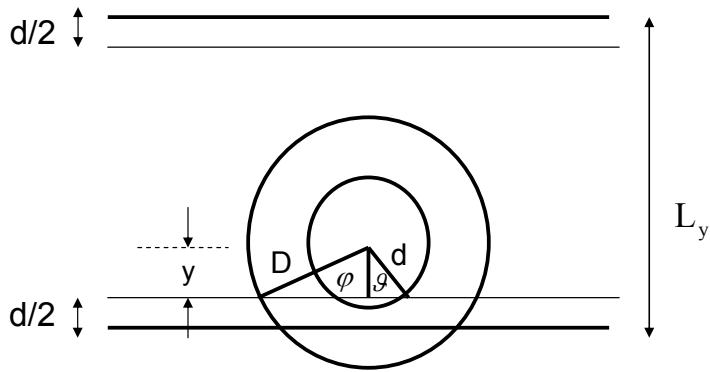}}
 \vspace{-6.5cm}
\caption{ The interaction disks of a particle with a two-step potential
in the case of
reflecting boundaries. The inner disk is of diameter $2d$ and that
of the outer disk is $2D$. The boundaries of the channel are indicated
by bold lines. The thin lines, are at a distance $d/2$ from the
respective boundary, and they represent the limits, which the centers
of the disks cannot cross. The vertical coordinate of the disk,
$y$, and the angles $\phi$ and $\vartheta$ are indicated.}
 \label{overlap_box3}
\end{figure}

\begin{figure}
\vspace{-6cm} \centering {
\includegraphics[width=12cm]{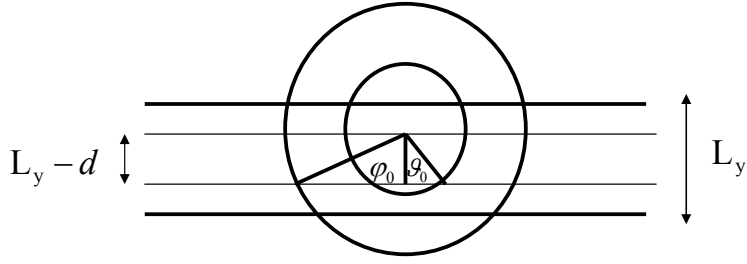}}
 \vspace{-6.5cm}
\caption{The interaction disks of a particle with a two-step potential in the case of
reflecting boundaries for $L_y<2d$, where the angles $\phi_0$ and
$\vartheta_0$ are defined. In the intermediate regime $2d<L_y<D+d$,
the smaller disk does not intersect the boundary of the channel, and
$\vartheta_0$ becomes $0$.} \label{overlap_box4}
\end{figure}

In calculating the overlap integral, one should distinguish between
three regimes.
\begin{itemize}
\item
For $L_y>D+d$ the integral may be expressed as (see Fig.
(\ref{overlap_box3}))
\begin{eqnarray}
q(L_y)&=& \pi (D^2-d^2) (e^{-\beta u}-1) -\pi d^2 \nonumber \\
&-& \frac{2}{L_y-d}\left[D^2 \int_0^Dh(\phi)dy -d^2\int_0^d
h(\vartheta)dy \right](e^{-\beta u}-1) \nonumber \\
&+& \frac{2}{L_y-d}d^2 \int_0^d h(\vartheta)dy ~,
\end{eqnarray}
where the function $h$ is given in Eq. (\ref{S-y}), and the angles
$\phi$ and $\vartheta$ are related to $y$ via (see Fig.
(\ref{overlap_box3}))
\begin{equation}
y=d \cos\vartheta = D \cos \phi ~.
\end{equation}
Evaluating the integrals, one obtains
\begin{equation}
q(L_y)=(\pi D^2 -\frac{4}{3} \frac{D^3}{L_y-d})(e^{-\beta u}-1) -
(\pi d^2 -\frac{4}{3} \frac{d^3}{L_y-d})e^{-\beta u} ~.
\label{q-reflecting_hard_soft>}
\end{equation}
\item
For $2d<L_y<D+d$ the overlap integral becomes
\begin{eqnarray}
q(L_y)&=& \pi (D^2-d^2) (e^{-\beta u}-1) -\pi d^2 \nonumber \\
&-& \frac{2}{L_y-d}\left[D^2 \int_0^{L_y-d}h(\phi)dy -d^2\int_0^d
h(\vartheta)dy \right](e^{-\beta u}-1) \nonumber \\
&+& \frac{2}{L_y-d}d^2 \int_0^d h(\vartheta)dy ~.
\end{eqnarray}
The integrals are readily evaluated to yield
\begin{eqnarray}
q(L_y)&=& (\pi D^2 -\frac{2D^3}{L_y-d} g(\phi_0))(e^{-\beta u}-1)
\nonumber \\
&-& (\pi d^2 -\frac{4}{3} \frac{d^3}{L_y-d})e^{-\beta u} ~.
\label{q-reflecting_hard_soft>}
\end{eqnarray}
where
\begin{equation}
g(\alpha)=\frac{2}{3} +\alpha\cos \alpha - \sin \alpha + \frac{1}{3}
\sin^3 \alpha ~,
\end{equation}
and (see Fig. (\ref{overlap_box4}))
\begin{equation}
\cos \phi_0 = \frac{L_y-d}{D} ~.
\label{phi0}
\end{equation}
\item
For $L_y < 2d$ the overlap integral is
\begin{eqnarray}
q(L_y)&=& \pi (D^2-d^2) (e^{-\beta u}-1) -\pi d^2 \nonumber \\
&-& \frac{2}{L_y-d}\left[D^2 \int_0^{L_y-d}h(\phi)dy
-d^2\int_0^{L_y-d}
h(\vartheta)dy \right](e^{-\beta u}-1) \nonumber \\
&+& \frac{2}{L_y-d}d^2 \int_0^d h(\vartheta)dy ~.
\end{eqnarray}
It yields
\begin{eqnarray}
q(L_y)&=& (\pi D^2 -\frac{2D^3}{L_y-d} g(\phi_0))(e^{-\beta u}-1)
\nonumber \\
&-& (\pi d^2 -\frac{2D^3}{L_y-d} g(\vartheta_0)) e^{-\beta u} ~,
\end{eqnarray}
where $\phi_0$ is given by Eq. (\ref{phi0}), and
\begin{equation}
\cos \vartheta_0 =\frac{L_y-d}{d} ~,
\end{equation}
(see Fig. (\ref{overlap_box4})).
\end{itemize}
The nature of the singularities of the pressure components at
$L_y=D+d$ and $L_y=2d$ can be analyzed as before. It is
straightforward to show that at both points the $P_{yy}$ components
exhibits a $3/2$ singularity, while the $P_{xx}$ component exhibits a
$5/2$ singularity.

\begin{figure}
\centering
{ \includegraphics[width=10cm]{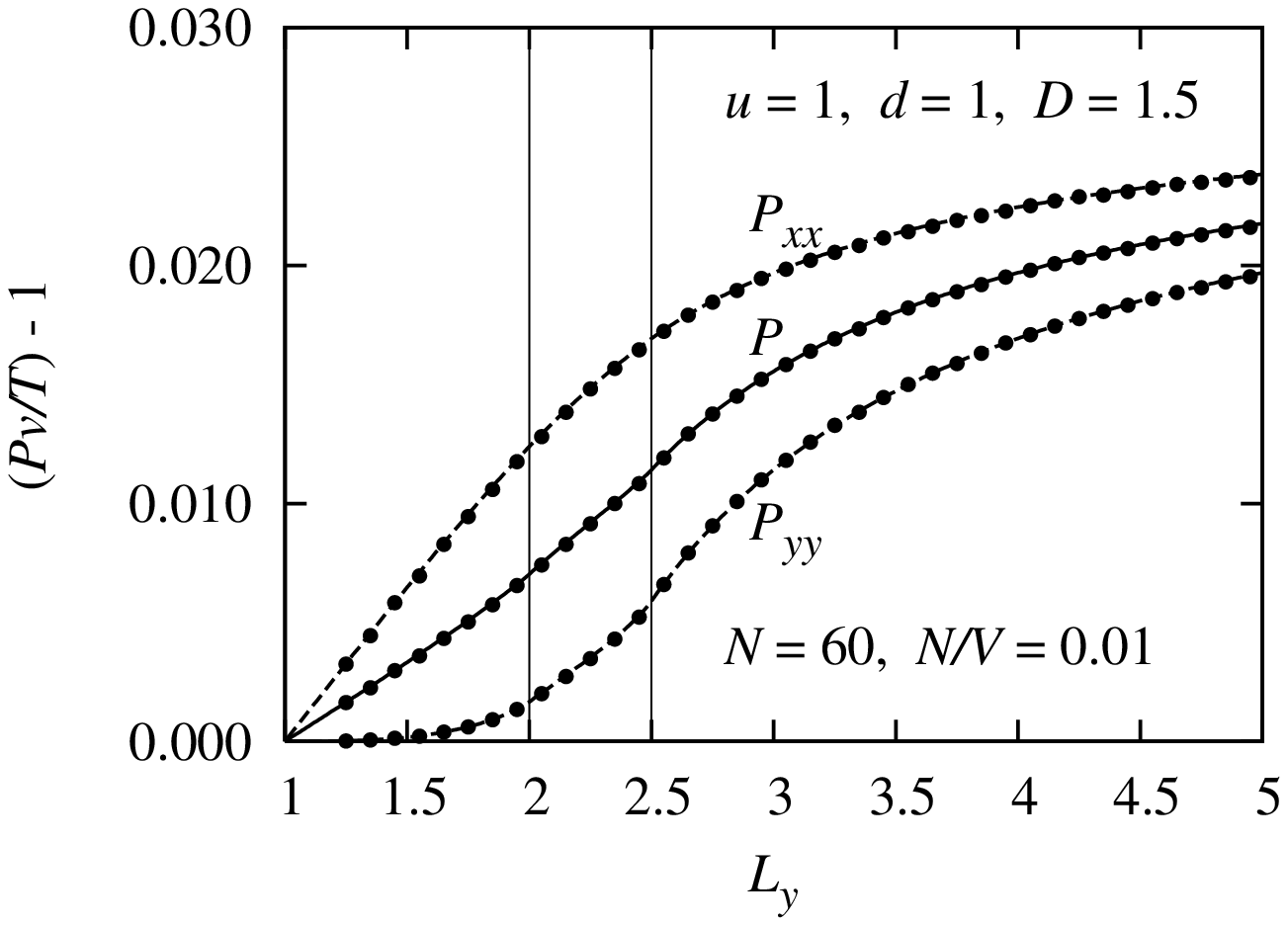}}
 \vspace{-3mm}
\caption{ Channel width dependence  of the potential-generated
pressures for the interaction potential (\ref{2step_potential}) with
$d=1$, $D=1.5$ and $u=1$. The dots are computer simulation results
for $N=60$ particles at a density $N/V = 0.01$. The smooth lines are
obtained from the theoretical overlap integrals of Sec.
\ref{soft_disks_hard_core}. The singular channel widths are
indicated by vertical lines. Reduced units are used, for which the
particle mass $m$ and the energy per particle, $E/N$, are unity.
 } \label{Fig15}
\end{figure}

\begin{figure}
\centering
{ \includegraphics[width=10cm]{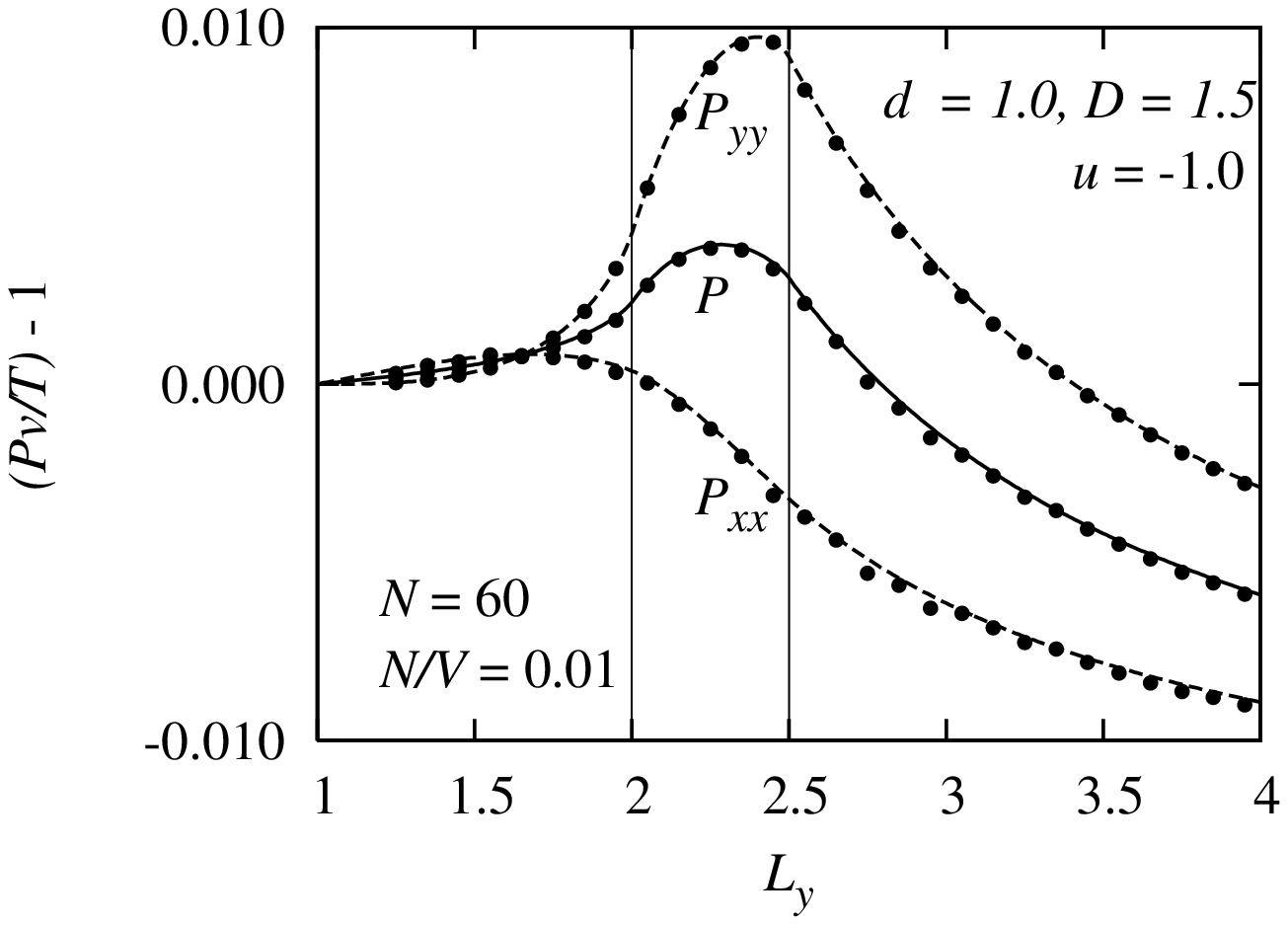}}
 \vspace{-3mm}
\caption{Channel width dependence  of the potential-generated
pressures for the interaction potential (\ref{2step_potential}) with
$d=1$, $D=1.5$ and $u=-1$. The dots are computer simulation results
for $N=60$ particles at a density $N/V = 0.01$. The smooth lines are
obtained from the theoretical overlap integrals of Sec.
\ref{soft_disks_hard_core}. The singular channel widths are
indicated by vertical lines. Reduced units are used, for which the
particle mass $m$ and the energy per particle, $E/N$, are unity. }
\label{Fig16}
\end{figure}
Comparisons of these results with computer simulations for $N = 60$
particles of equal mass $m$ are provided in Fig. \ref{Fig15} for the
potential (\ref{2step_potential}) with a positive step $u = 1$, and
in Fig. \ref{Fig16} for the case of a negative step potential, $u =
-1 $. All quantities are given in reduced units, for which the
particle mass $m$, the hard core diameter, $d$, and the total energy
per particle, $E_0 = E/N$, are unity. The outer diameter is taken to
be $D=1.5$. The singular points at $L_y = d + D = 2.5$ and $L_y = 2d
= 2$ are marked by the vertical lines. The density $N/V = 0.01$. For
the computation of the theoretical pressures resulting in the smooth
lines of Figs. \ref{Fig15} and \ref{Fig16}, the slight variation of
the kinetic energy and, hence, of the temperature with the channel
width  was taken into account. The agreement between the theoretical
expressions and the computer simulation results for the potential
part of the pressures is very satisfactory.
%
\section{Summary}
\label{conclusions}
%

In the paper we studied the pressure tensor of a system of disks
moving in a narrow two dimensional channel, with either periodic or
reflecting boundary conditions. We considered the low density regime
using the Mayer cluster expansion, and tested the validity of the
expansion using molecular dynamics studies. It is found that
whenever the two-body interaction potential between disks, $u(r)$,
exhibits a singularity at some distance $r_0$, the pressure tensor
exhibits a singularity as a function of the channel width, at one or
more widths which are simply related to $r_0$. By studying several
classes of interaction potentials, some rather general conclusions
regarding the singularities of the pressure tensor can be reached.

In the case of periodic boundary conditions, singularities take
place at channel widths $L_y = 2 r_0/n$ with $n=1,2, \dots$~. For
potentials which exhibit a discontinuity at $r_0$, the transverse
pressure, $P_{yy}$, exhibits a $1/2$ singularity while the
longitudinal component, $P_{xx}$, exhibits a weaker $3/2$ singularity.
For potentials which are continuous at $r_0$, and whose singular
part vanishes as $(r_0-r)^\kappa$, the transverse pressure exhibits
a $2\kappa +1/2$ singularity while the singularity of the
longitudinal pressure is $2\kappa +3/2$. Although these results have
been demonstrated for specific interaction potentials $u(r)$, they
are rather general, as they are related only to the nature of the
singularity of the potential.

In the case of reflecting boundary conditions the pressure tensor
exhibits a singularity at $L_y=r_0$. The singularity is weaker than
that of the case of periodic boundary conditions. Particularly, it
was found that for a potential which is discontinuous at $r_0$~, the
transverse component of the pressure exhibits a $3/2$ singularity,
while the longitudinal component exhibits a weaker $5/2$
singularity.

%
\section{Appendix}
\label{Appendix}
%

There are a few minor misprints in some of the equations in the
paper on hard disks \cite{FMP04} which are corrected below. These
misprints do not affect any of the expressions for the pressure
derived in that paper, or any of the numerical results.

\noindent In particular, Equation (3) of Ref. \cite{FMP04}
should read
\begin{equation}
s \equiv \frac{S}{N}=  \ln \left(v-\frac{q(L_y)}{2}\right), \nonumber
\label{entropycorr}
\end{equation}
and Equation (13) in \cite{FMP04} should become
\begin{eqnarray}
\frac{P_{xx}v}{kT} & = & L_x \left(\frac{\partial s}{\partial
L_x}\right)_{L_y} ,\nonumber \\
\frac{P_{yy}v}{kT} & = & L_y \left(\frac{\partial s}{\partial
L_y}\right)_{L_x} .  \nonumber
\end{eqnarray}
Another misprint concerns the definition of ${\bf  v}_c$ in the
expression for the virial in Eq. (8).  ${\bf  v}_c$ is the velocity change
of a particle $i$ taking part in a binary collision $c$.

 \section{Acknowledgments}
HAP is grateful for the hospitality accorded to him at the Weizmann
Institute of Science. Support from the Austrian Science Foundation
(FWF), grant P18798, the Minerva Foundation with funding from the
Federal German Ministry for Education and Research and the Albert
Einstein Minerva Center for Theoretical Physics is gratefully
acknowledged. A part of this work was carried out at the Erwin
Schr\"odinger Institute in Vienna on the occasion of a workshop in
June 2008.

\end{document}